\documentclass[preprint,12pt]{elsarticle}

\usepackage{amssymb}
\usepackage{amsthm}

\usepackage{color}

\begin{document}

\begin{frontmatter}



\title{A frequency portrait of Low Earth Orbits}


\author[au1]{Giulia Schettino\corref{cor1}} 
\ead{g.schettino@ifac.cnr.it}
\author[au1]{Elisa Maria Alessi}
\ead{em.alessi@ifac.cnr.it}
\author[au1]{Alessandro Rossi} 
\ead{a.rossi@ifac.cnr.it}
\author[au1,au2]{Giovanni B. Valsecchi}
\ead{giovanni@iaps.inaf.it}

\cortext[cor1]{Corresponding author}
\address[au1]{IFAC-CNR, Via Madonna del Piano 10, 50019 Sesto Fiorentino (FI) - Italy}
\address[au2]{IAPS-INAF, Via Fosso dei Cavalieri 100, 00133 Rome - Italy}

\begin{abstract}
  In this work we deepen and complement the analysis on the dynamics
  of Low Earth Orbits (LEO), carried out by the authors within the
  H2020 ReDSHIFT project, by characterising the evolution of the
  eccentricity of a large set of orbits in terms of the main frequency
  components. Decomposing the quasi-periodic time series of
  eccentricity of a given orbit by means of a numerical computation of
  Fourier transform, we link each frequency signature to the dynamical
  perturbation which originated it in order to build a frequency chart
  of the LEO region. We analyse and compare the effects on the
  eccentricity due to solar radiation pressure, lunisolar
  perturbations and high degree zonal harmonics of the geopotential
  both in the time and frequency domains.  In particular, we identify
  the frequency signatures due to the dynamical resonances found in
  LEO and we discuss the opportunity to exploit the corresponding
  growth of eccentricity in order to outline decommissioning
  strategies.
\end{abstract}

\begin{keyword}
LEO, frequency analysis, SRP, lunisolar perturbations
\end{keyword}

\end{frontmatter}



\section{Introduction}

It is known that the proliferation of space debris in the Low Earth
Orbit (LEO) region has already become a critical issue to handle. In
this context, as part of the H2020 ReDSHIFT (Revolutionary Design of
Spacecraft through Holistic Integration of Future Technologies)
project \cite{RossiSD,RossiAero}, a deep analysis to search for
passive deorbiting solutions in LEO was carried out, by performing an
accurate mapping of the phase space in order to identify stable and
unstable regions. A detailed description of the results of the LEO
cartography was presented by the authors in \cite{AlessiCMDA}, while
in \cite{AlessiMNRAS} a general analysis on the role that resonances
induced by solar radiation pressure (SRP) can play in assisting the
deorbiting was provided.

In general, the key idea investigated in those works was to identify
the orbits, and the associated mechanisms, where dynamical
perturbations can induce a significant growth of the orbital
eccentricity, in order to facilitate passive disposal. Accordingly
with our findings, we concluded that in the case of a typical intact
object in LEO, with an area-to-mass ratio of the order of
$A/m=10^{-2}$ m$^2/$kg, perturbations as SRP, lunisolar effects and
high degree zonal harmonics cannot ensure the reentry on their own but
only in combination with the atmospheric drag. If the spacecraft is,
instead, equipped with an area augmentation device, which increases
the effective $A/m$ by, e.g., two orders of magnitude, then we
concluded that SRP alone can drive the dynamics, if the initial
inclination of the orbit is close enough to a resonant inclination,
given semi-major axis and eccentricity.

In the present work, we make a deeper analysis of the role of
resonances which act in the LEO dynamics, by
characterising the eccentricity of a set of orbits in terms of
periodic components. Starting from the quasi-periodic time series of
eccentricity, computed for a dense grid of initial conditions, we
decompose the series in the main spectral components by means of a
numerical computation of the Fourier transform. Then, we link each
frequency component with the dynamical perturbation responsible for
that signature. In this way, we have an additional tool to explore the
relative importance of each given gravitational or non-gravitational
perturbation in LEO as a function of the initial orbital elements. The
final goal of such analysis is to support the cartography in
identifying the orbits where a significant growth of eccentricity, led
by one or more perturbations, can assist the passive disposal of
objects at their end-of-life. The same analysis can also serve to
identify the periodic drifts that
operational orbits could experience.

In the past, the study of the chaotic dynamics within the Solar System
led \cite{Laskar1990} to devise a method for a numerical estimation
of the size of the chaotic zones, based on the variation in time of
the main frequencies of the system. Since then, the algorithm for the
frequency analysis was developed to study the stability of the orbits
in many multi-dimensional conservative systems, in order to provide a
global representation of the dynamics
\cite{Laskar1992,Laskar1993b}. The Frequency Map Analysis algorithm
(Numerical Analysis of Fundamental Frequencies - NAFF) is based on a
refined and iterative numerical search for a quasi-periodic Fourier
approximation of the solution of the system over a finite time span
\cite{Laskar1993a}. Considering, in particular, the issue of optimal
design of artificial satellite survey missions around a
non-axisymmetric body, Noullez et al. \cite{Noullez} proposed an alternative method
with respect to the standard Fourier transform approach to
characterise satellite orbits by computing the periodic components in
order to identify regular orbits, meant as orbits whose inclination
and eccentricity do not vary significantly over a given time scale.
Concerning in particular the LEO region, Celletti and Gale\c{s} \cite{Cell2017} studied the
dynamics of resonances in LEO with the aim of identifying the location
of equilibrium position and their stability. Within the common scope
of defining suitable post-mission disposal orbits, they studied
analytically, by means of a toy-model, whether an object is located in
a stable or chaotic region. In such a way, the identification of
stable orbits in LEO suggests the detection of possible graveyard
orbits. In this paper, we focus on the possibility of exploiting the
eccentricity growth induced by one or more dynamical perturbations at
given orbits to facilitate the end-of-life reentry and we deepen this
analysis by characterising the eccentricity evolution in terms of its
main frequency components. A comprehensive characterization of the
dynamical evolution of the eccentricity is a key ingredient in order
to identify, among other things, possible disposal strategies for
operational and future spacecraft.

The paper is organised as follows: in Section~\ref{sec:mm} we
briefly describe the dynamical model adopted for the numerical
propagation and we introduce the method to identify the frequency
signatures which characterise the eccentricity evolution of a set of LEO
orbits. In Section~\ref{sec:res} we outline the results of our
analysis, comparing the results of numerical propagation in the
time domain with the findings of the frequency
characterisation. Finally, in Section~\ref{sec:concl} we draw some
conclusions.

\section{Dynamical model and methods}
\label{sec:mm}

As mentioned before, within the scope of ReDSHIFT, we performed an
extensive mapping of the LEO phase space by propagating more than 3
million orbits, as described in
\cite{AlessiSD,AlessiIAC,AlessiCMDA}\footnote{All the papers related
  to the project are available on the ReDSHIFT website at
  http://redshift-h2020.eu/documents/.}, spanning from $500\,$km to
$3000\,$km of altitude over the Earth surface, considering a wide
range of eccentricities, from 0 up to 0.28, and inclinations, from
$0^{\circ}$ to $120^{\circ}$, 16 different $(\Omega, \omega)$
configurations and two initial epochs. In the following, we limit our
analysis to the case of right ascension of the ascending node,
$\Omega$, and argument of perigee, $\omega$, both equal to
$0^{\circ}$, with the initial epoch set to 21 June 2020. The orbital
propagation was carried out over a time span of 120 years by means of
the semi-analytical orbital propagator FOP (Fast Orbit Propagator, see
\cite{Ans,Rossi09} for details), which accounts for the effects of
$5\times 5$ geopotential, SRP (assuming the cannonball model),
lunisolar perturbations and atmospheric drag (below 1500 km of
altitude). Two possible values of the area-to-mass ratio were
considered: $A/m=0.012$ m$^2/$kg, selected as a reference value for
typical intact objects in LEO, and $A/m=1$ m$^2/$kg, a representative
value for a small satellite equipped with an area augmentation device,
as a solar sail \cite{ColomboIAC}. More details on the adopted model
can be found in \cite{AlessiCMDA}. The results of the cartography can
be displayed in contour maps showing the lifetime or the maximum
eccentricity over the propagation interval as a function of the
initial inclination and eccentricity, for each initial semi-major
axis. A large set of maps can be found on the ReDSHIFT
website\footnote{http://redshift-h2020.eu/results/leo .}. In the
following Sections, some examples will be provided.


\subsection{Dynamics in the time domain}
\label{sec:model}

We are particularly interested in studying the time
evolution of the eccentricity. Indeed, within the search for passive
disposal solutions in LEO, the identification of orbits which can
experience a significant growth of eccentricity becomes crucial, since
in this case the lowering of the orbital perigee helps drag in being
effective. Moreover, a variation in eccentricity causes an altitude
variation which could become an issue also at the operational stage,
for instance in the case of a large constellation.

Lagrange planetary equations (e.g., \cite{Roy}) show that SRP,
lunisolar perturbations and high degree zonal harmonics\footnote{The
  oblateness of the Earth, $J_2$, does not affect the evolution of the
  eccentricity over long term (e.g. \cite{Roy}).} cause long term
periodic variations in the evolution of eccentricity, which become
quasi-secular in the vicinity of a resonance involving the rate of the
right ascension of the ascending node, $\Omega$, and the argument of
perigee, $\omega$. In particular, we can write the instantaneous
variation of $e$ due to a given perturbation in the general form:
\begin{equation}
\frac{de}{dt}=T(a,e,i)\,\sin\psi(\Omega,\omega,\lambda_S)\,,
\label{dedt}
\end{equation}
where $T$ is a coefficient which depends on $(a,e,i)$ according to the
given perturbation and the argument $\psi$ can be written in general
terms as:
\begin{equation}\label{eq:psi}
\psi =\alpha\Omega +\beta\omega +\gamma\lambda_S\,,
\end{equation}
where $\alpha,\,\beta,\,\gamma=0,\pm 1,\pm 2$ depending on the
perturbation and $\lambda_S$ is the longitude of the Sun with respect
to the ecliptic plane, set as $\lambda_S=90.086^{\circ}$ at the
starting epoch. A resonance occurs when the condition
$\dot\psi\simeq 0$ is satisfied.

\begin{table}
	\centering
	\caption{List of the main resonances expected to be found in LEO: argument $\psi_j$, values of the coefficients $\alpha,\,\beta,\,\gamma$ and corresponding index $j$. Resonances from $j=1$ to $j=6$ are due to SRP; resonances 7 and 8 are singly averaged solar gravitational resonances; resonances from 9 to 11 are doubly averaged lunisolar resonances.}
	\label{tab:res}
	\begin{tabular}{lrrrc} 
		\hline
		Argument $\psi_j$ & $\alpha$ & $\beta$ & $\gamma$ & index $j$\\
		\hline
		$\Omega +\omega -\lambda_S$ & 1 & 1 & $-1$ & 1 \\
		$\Omega -\omega -\lambda_S$ & 1 & $-1$ & $-1$ & 2 \\
		$\omega -\lambda_S$ & 0 & 1 & $-1$ & 3 \\
                $\omega +\lambda_S$ & 0 & 1 & 1 & 4 \\
                $\Omega +\omega +\lambda_S$ & 1 & 1 & 1 & 5 \\
                $\Omega -\omega +\lambda_S$ & 1 & $-1$ & 1 & 6 \\
		\hline
                $\Omega+2\omega-2\lambda_S$ & 1 & 2 & $-2$ & 7 \\
                $2\Omega +2\omega -2\lambda_S$ & 2 & 2 & $-2$ & 8 \\
                \hline
                $\omega$ & 0 & 1 & 0 & 9 \\
                $\Omega +2\omega$ & 1 & 2 & 0 & 10 \\
                $2\Omega +2\omega$ & 2 & 2 & 0 & 11 \\
                \hline
	\end{tabular}
\end{table}

The list of the resonances expected from the theory and found by means of the
LEO cartography \cite{AlessiCMDA} are shown in Table \ref{tab:res},
where the corresponding expression for $\psi$ and the value of
$\alpha,\,\beta,\,\gamma$ are highlighted, together with an index
($j=1,..11$) associated to each resonance. Resonances indexed from 1
to 6 correspond to the condition
\begin{equation}\label{eq:dotpsi}
\dot\psi=\bar{\alpha}\dot\Omega\pm\dot\omega\pm\dot\lambda_S\simeq
0\,,
\end{equation}
with $\bar{\alpha}=0,1$, and are associated to the zero-order
expansion of the SRP disturbing function (e.g.,
\cite{Hughes77,Krivov}). Resonances 7 and 8 are singly averaged solar
gravitational resonances (e.g., \cite{Hughes80,Breiter}), while
resonances from 9 to 11 are associated to doubly averaged lunisolar
gravitational perturbations (e.g., \cite{Hughes80}). The rate of
$\Omega$ and $\omega$ can be found by applying the Lagrange planetary
equations and accounting in principle for both the effects of $J_2$
and SRP, while the effect of lunisolar perturbations can be neglected
(e.g., \cite{Milani87}). The explicit expressions have been given,
for instance, in \cite{AlessiMNRAS}. In practice, in
\cite{AlessiMNRAS} we have shown that, for an initial orbit with
$\Omega=\omega=0^{\circ}$, at the assumed initial epoch (which
corresponds to $\lambda_S\approx 90^{\circ}$), the rate of $\Omega$
and $\omega$ due to SRP vanishes.

\begin{figure}
\centering
	\includegraphics[width=0.8\textwidth]{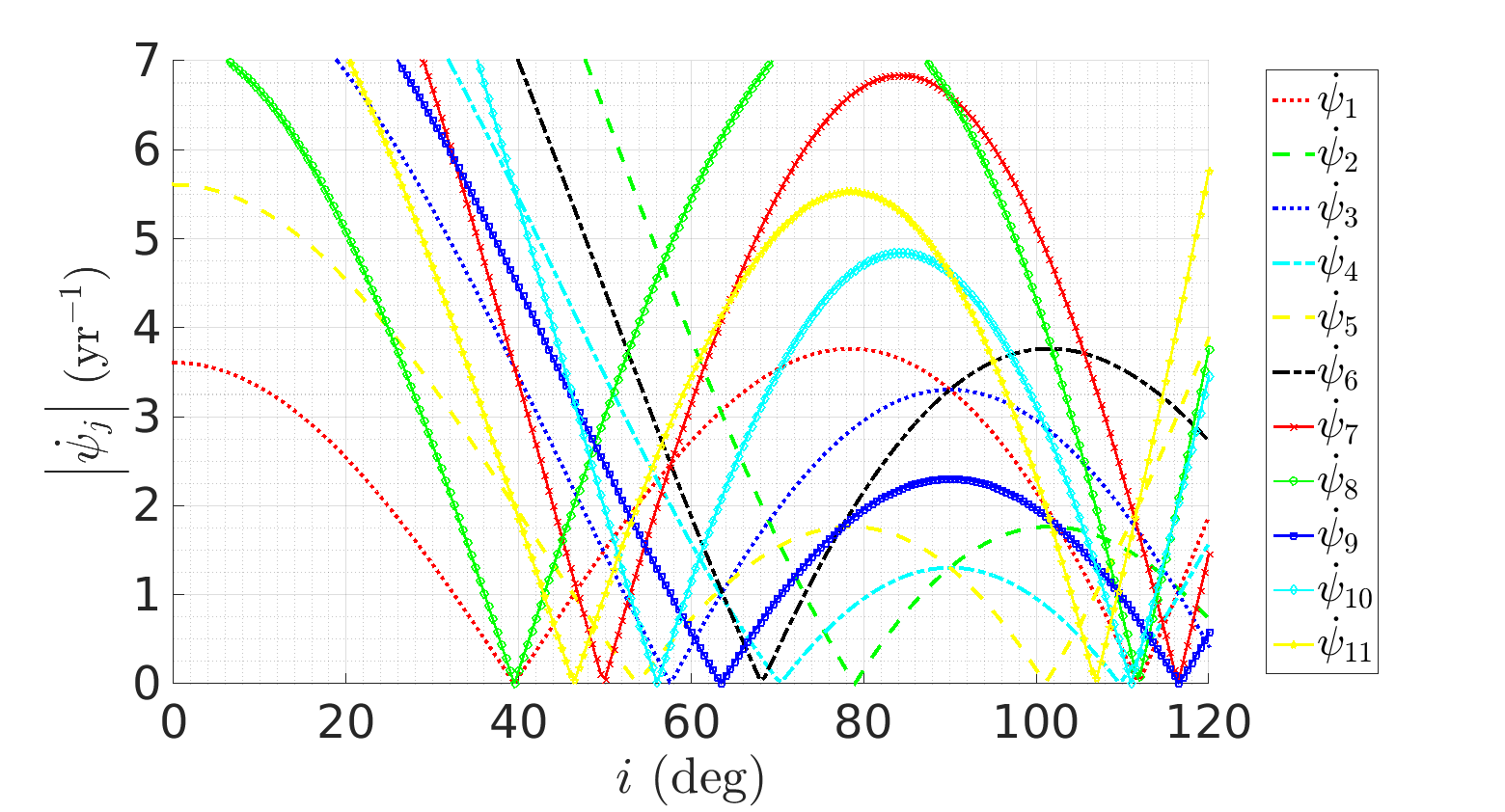}
        \caption{Behaviour of $|\dot\psi|$ for each perturbing term
          $j=1,..11$ as a function of the inclination, for $e=0.001$
          and $a=7978$ km, in the case $A/m=1$ m$^2/kg$.}
    \label{fig:Fig1}
\end{figure}

In Figure~\ref{fig:Fig1} we display the behaviour of $|\dot\psi_j|$
for each perturbing term highlighted in Table~\ref{tab:res}
($j=1,..11$) as a function of the inclination
$i\in [0^{\circ}:120^{\circ}]$ for the case of a quasi-circular orbit
($e=0.001$) with a semi-major axis $a=7978$ km.  The figure shows that
curves associated to different perturbations may intersect and overlap
creating a dense network of resonances in the phase space. Thus,
depending on the given inclination, it may be hard to distinguish
between the concurrent effect of different perturbations and to link
the dynamical effect to the perturbation which produces it. To
overcome this problem, we can take advantage of the fact that the
adopted orbital propagator is set up in such a way that each dynamical
perturbation in the model can be individually turned on or off. Since
we aim at identifying the specific effect of a given perturbation on
the eccentricity evolution and at characterising it in the frequency
domain, we consider two simplified models, which fit our purposes:
\begin{itemize}
\item model I: SRP on; lunisolar
  perturbations and drag off; geopotential: only $J_2$;
\item model II: SRP off; lunisolar perturbations and drag on;
  geopotential: $5\times 5$.
\end{itemize} 

Model I is particularly suitable to study the SRP effects on the
eccentricity in the case of high $A/m$ objects, when only SRP and drag
play a primary role in the evolution. In the case of $A/m=1$ m$^2/$kg,
atmospheric drag is effective in driving a reentry within 25 years for
pericenter altitudes up to $1050\,$km (see
\cite{AlessiCMDA,Schettino2018}). Since this is a relatively high
value, in order to focus on the effect due to SRP, we have decided to
switch off the perturbation due to the atmospheric drag.

Model II, instead, is appropriate to study the effects led by
lunisolar perturbations and high degree zonal harmonics: removing from
the model the presence of SRP, we avoid the chance of mismodelling,
since the resonant inclinations corresponding to lunisolar
perturbations and geopotential can be close to those associated with
SRP, as appears from Figure~\ref{fig:Fig1}. In this case, adopting the
low or the high value of $A/m$ does not affect the eccentricity
evolution.

\subsection{Frequency characterisation of the eccentricity}
\label{sec:fc}

The starting point for the frequency characterisation is to process
the discrete eccentricity time series of a given initial orbit to
obtain the discrete Fourier transform through a standard Fast Fourier
Transform (FFT) algorithm (e.g., \cite{Opp}), based on the
Cooley-Tukey algorithm \cite{Cooley}. The basic idea is to identify
the frequency and the amplitude of the main spectral features in the
frequency series. The criterion we adopt is to account for any
signature whose amplitude is, at least, 10 times stronger than the mean
value of the spectrum in the surrounding area.

A first issue to be considered concerns the time sampling
$\Delta t$ of the input series to be transformed. Indeed, the sampling
frequency is $f_s=1/\Delta t$ and, from Nyquist theorem, it
follows that $f_s/2$ is the highest frequency we can capture from our
analysis. Since the perturbations we are interested in have
periodicity of the order of months to years\footnote{We recall that
  moving close to a resonant orbit, the period of the perturbation
  acting on the eccentricity becomes gradually longer, up to
  quasi-secular if the orbital inclination corresponds exactly to a
  resonant condition.}, the sampling $\Delta t=1\,$day, adopted in
\cite{AlessiSD,AlessiIAC,AlessiCMDA}, is fully reasonable. On the
other side, a more critical issue involves the lowest detectable
frequency by our analysis, which is limited by $2/T$, where $T$ is the
duration of the time series. This means that with the adopted time
span of 120 years, signatures with periodicity up to 60 years
would be, in principle, identified. In practice, signatures due to
perturbations with periodicity of more than some years are poorly
sampled by definition. Thus, we propagate the set of orbits of
interest for a longer time span, 600 years, in order to catch
unambiguously signatures with periodicity of some tens of years, as
expected in the vicinity of a resonance.

\section{Analysis of the numerical results}
\label{sec:res}

The general results of the LEO dynamical mapping was already
extensively described in \cite{AlessiCMDA}.  In the following, we
present the results obtained by assuming the two simplified dynamical
models, described in Section~\ref{sec:model}. First, we consider the
case of model I, i.e., we focus on the effect of SRP in the case of
the augmented $A/m$ ratio: we briefly recall the main findings in
terms of time evolution of the eccentricity, then we discuss the
results of the characterisation in terms of frequency
components. Next, we present the same analysis in the case of model
II, focusing on the effects of lunisolar perturbations and high degree
zonal harmonics.

\subsection{Model I}

\subsubsection{Analysis in the time domain}

\begin{figure}
  \begin{center}   
  \includegraphics[width=0.48\textwidth]{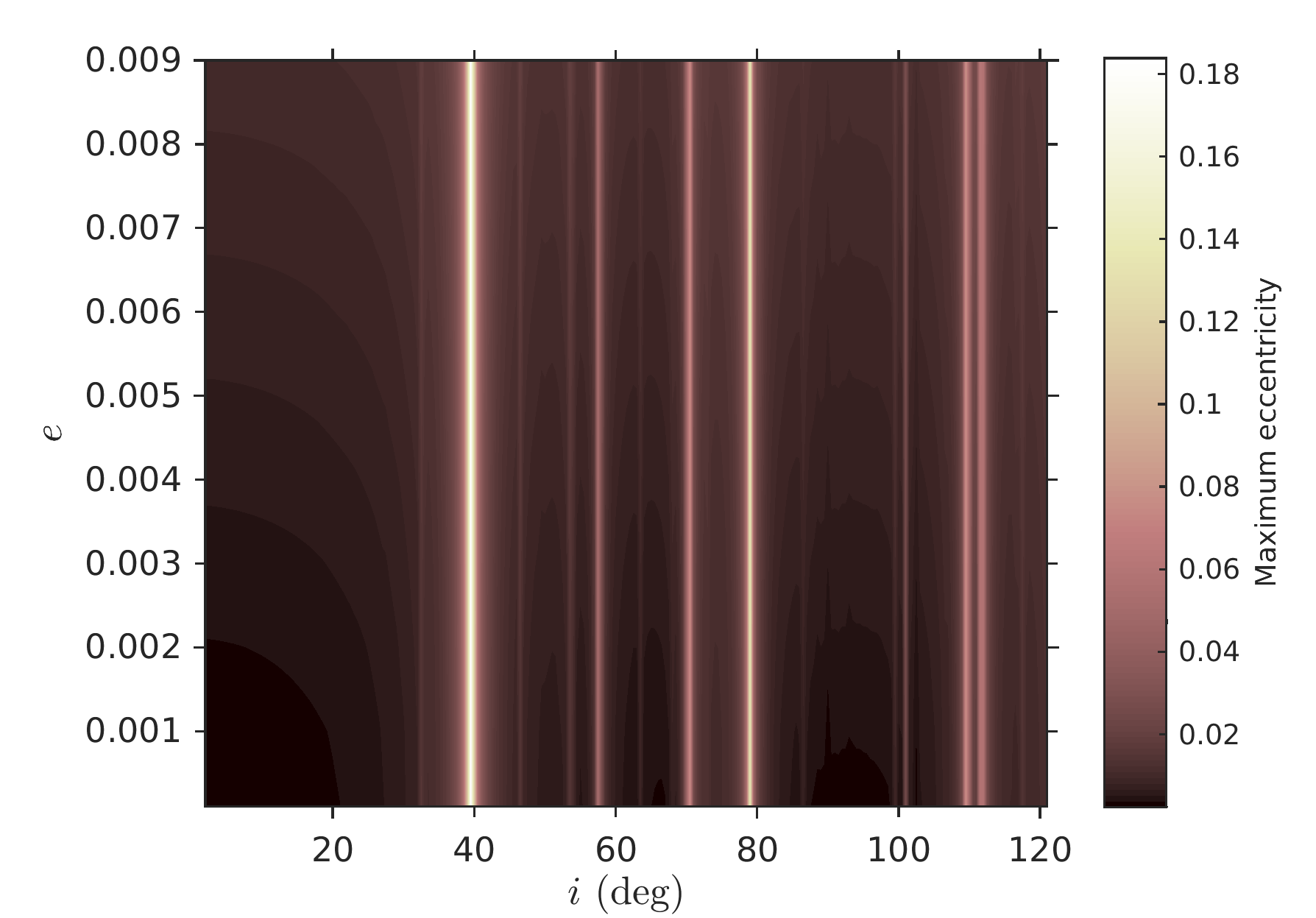}
       \includegraphics[width=0.48\textwidth]{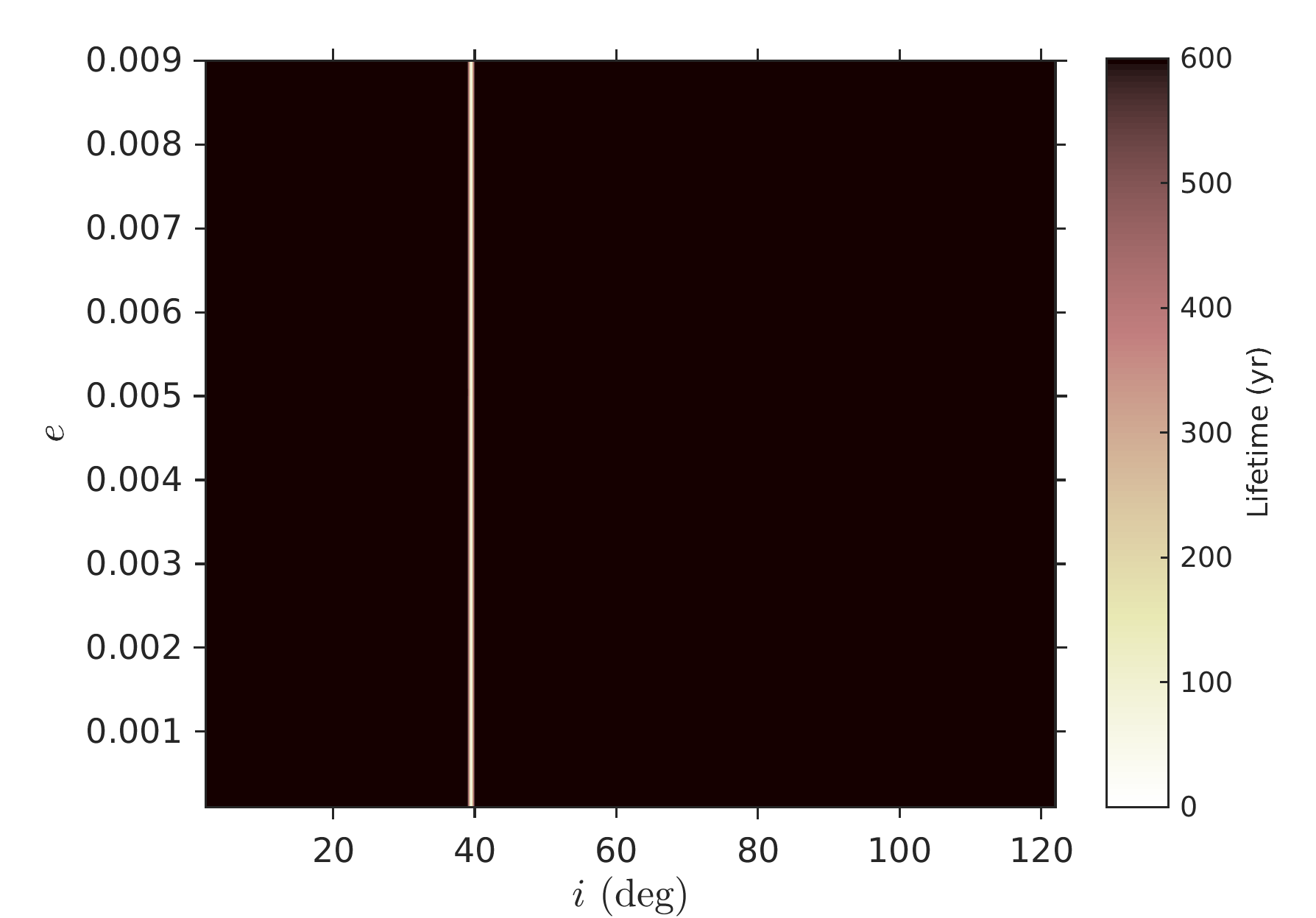}
  \includegraphics[width=0.48\textwidth]{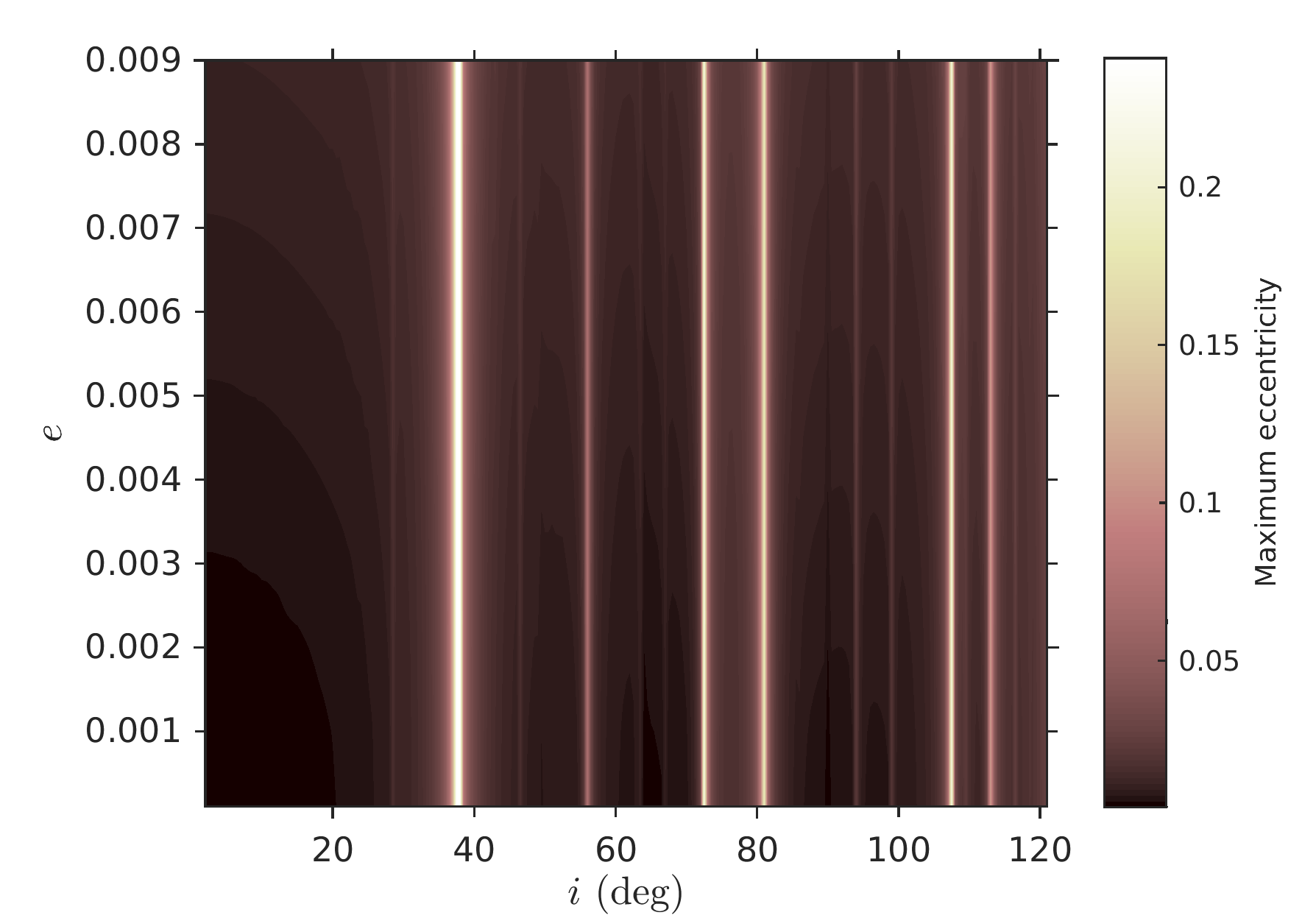}
       \includegraphics[width=0.48\textwidth]{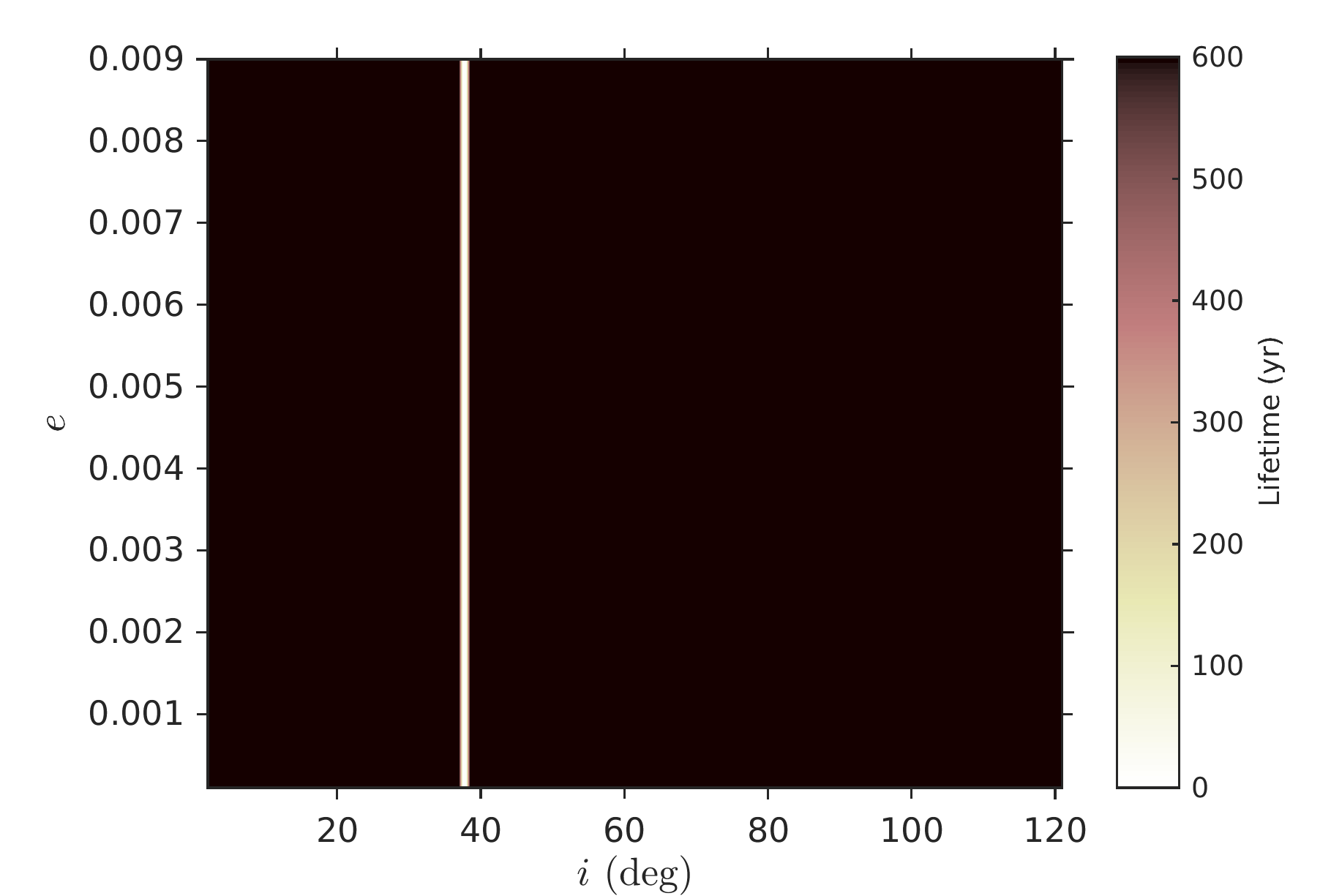}
 \caption{Maximum eccentricity (left column) and lifetime over 600
   years (right column; in the color bar: in years) as a function of
   the initial inclination at steps of $\Delta i=0.5^{\circ}$ and
   $e$ at steps of $\Delta e=0.001$, assuming model I and $A/m=1$ m$^2/$kg,
   for the initial orbits at $a=7978\,$km (top) and $a=8578\,$km
   (bottom), with $\Omega=\omega=0^{\circ}$ and initial epoch 21 June 2020.}
      \label{contour1} 
\end{center}
\end{figure}

We recall that the model accounts, in this case, only for the effect
of SRP and $J_2$, while drag and lunisolar perturbations are turned
off. We propagate the orbits assuming $A/m=1\,$m$^2/$kg and we look
for the inclinations where a growth of eccentricity due to SRP
occurs. Some illustrative results are shown in Figure~\ref{contour1}:
on the left we show the maximum eccentricity achieved over 600 years
of propagation as a function of the initial inclination and
eccentricity, for initial $a=7978$ km (top) and $a=8578$ km (bottom),
respectively. On the right panels we display the corresponding
lifetime, in years.  We recall that the atmospheric drag is effective
up to 1050 km of altitude for the adopted $A/m$ ratio. Thus, we
selected on purpose two reference values for the initial semi-major
axis which are significantly above the region where drag plays a
role. If the effect of SRP is able to lower the perigee below 1050 km,
then the removal of the drag from the model allows to check if the
chance to reenter or not can be ascribed solely to SRP.

The lifetime panels show that, in the case an area augmentation device
is available on-board, even for high altitude quasi-circular orbits, a
reentry driven by SRP alone is feasible for inclinations in the
vicinity of $40^{\circ}$, which corresponds to the resonant condition:
\begin{equation}
\dot\psi_1=\dot\Omega+\dot\omega-\do\lambda_S\simeq 0\,.
\end{equation}
In the case of initial $a=7978$ km, reentry can be achieved in about 7
years for initial $e$ ranging from 0.0001 to 0.009 thanks to SRP
alone, for an initial orbit at $i=39.5^{\circ}$. In the case of
initial $a=8578$ km, SRP allows to reenter within 10 years at initial
$i=37.5^{\circ}$ and in about 16 years for $i=38^{\circ}$.  The other
resonances due to SRP, although not able to drive a reentry, cause,
anyway, a remarkable growth in eccentricity, as can be seen from the
left panels of Figure~\ref{contour1}, which can be exploited to lower
the perigee of the orbit. Referring also to Figure~1 in
\cite{AlessiMNRAS}, which shows the location of the 6 main SRP
resonances as a function of $i$ and $a$, we can identify the following
resonances corresponding to the bright inclination ``corridors'':
\begin{itemize}
\item $\dot\psi_1\simeq 0$ around $i=40^{\circ}$ (and
  $i=113^{\circ}$);
\item $\dot\psi_2\simeq 0$ around $i=80^{\circ}$;
\item $\dot\psi_3\simeq 0$ and $\dot\psi_5\simeq 0$ around
  $i=58^{\circ}$ and $i=54^{\circ}$, respectively, in the top panel
  ($a=7978$ km), while they intersect around $i=56^{\circ}$ at $a=8578$
  km;
\item $\dot\psi_4\simeq 0$ and $\dot\psi_6\simeq 0$, both occurring
  in the vicinity $i=70^{\circ}$.
\end{itemize}

Moreover, we can recognise other features at specific inclinations,
appearing as fainter, but still visible, signatures. They can be
associated to higher-order terms in the expansion of the SRP
disturbing function (e.g., \cite{Hughes77}): in
Section~\ref{sec:3.modI} their identification will be assisted by the
analysis in terms of frequencies.

\begin{figure}
\centering
\includegraphics[width=0.7\textwidth]{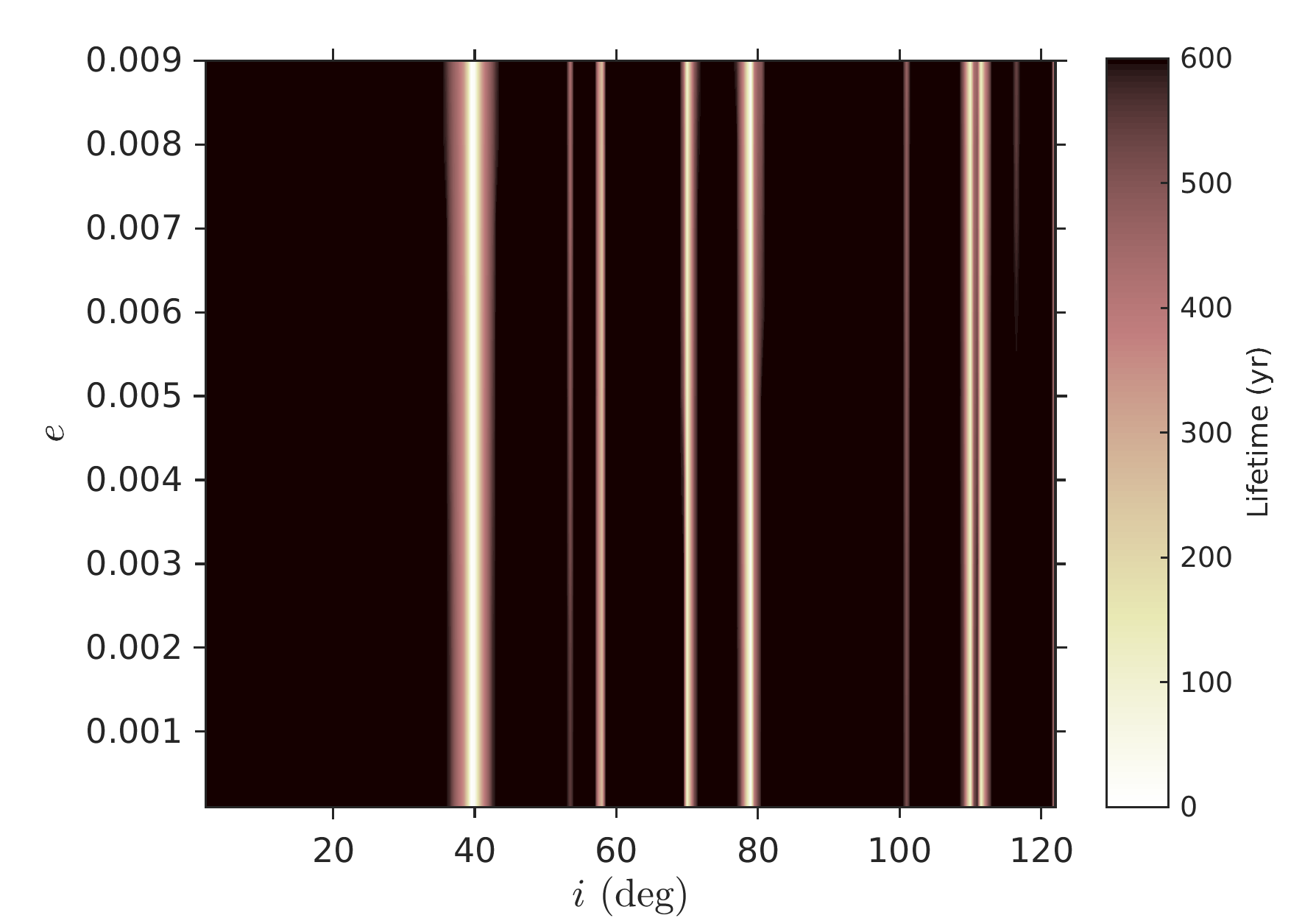}
        \caption{Lifetime (in the color bar: in years)
          as a function of the initial inclination at steps of
          $\Delta i=0.5^{\circ}$ and $e$ at steps of $\Delta e=0.001$,
          assuming model I with atmospheric drag and $A/m=1$ m$^2/$kg,
          for the initial orbit at $a=7978\,$km, with
          $\Omega=\omega=0^{\circ}$ and initial epoch 21 June 2020.}
    \label{contourSRP2}
\end{figure}

For completeness, turning on the contribution due to the atmospheric
drag in the model, we find that the synergic effect of SRP and drag
can support reentry also at different values of inclinations
(resonances) but, typically, only over long time scales. This is shown
in Figure~\ref{contourSRP2}, in the case of an initial orbit at
$a=7978$ km and $e=0.001$, assuming now model I with the further
contribution of the drag. For the same initial orbit,
Table~\ref{tab:life} shows the lifetime associated to the initial
inclination corresponding to the six SRP resonances. The table points
out that the addition of the drag in the model can assist the reentry
at inclinations close to the resonant ones, but only in the case of
resonance 2 (in addition to resonance 1) the reentry can take place in
less than 25 years.
\begin{table}
	\centering
	\caption{Resonant inclination $i_{res}$ and lifetime (in years) for each of the six main SRP resonances, in the case of initial $a=7978$ km and $e=0.001$, assuming model I with the addition of atmospheric drag.}
	\label{tab:life}
	\begin{tabular}{ccc} 
		\hline
		Resonance & $i_{res}$ & Lifetime (yr)\\
		\hline
		1 & $39.5^{\circ}$ & 6.5 \\
		2 & $79.0^{\circ}$ & 13.5 \\
                3 & $58.0^{\circ}$  & 261  \\
                4,6 & $70.0^{\circ}$ & 99  \\
                5 & $53.5^{\circ}$ & 545  \\
                \hline
	\end{tabular}
\end{table} 

\subsubsection{Analysis in the frequency domain}
\label{sec:3.modI}

The analysis of the maximum eccentricity maps (Figure~\ref{contour1} -
left panels) shows that, in addition to the six resonances due to the
zero-order expansion of the SRP disturbing function, other fainter
signatures can be observed at given inclinations. Thus, to build a
complete picture of the eccentricity evolution in the LEO phase space
we need to include the first-order terms in the expansion of the SRP
disturbing function (e.g., \cite{Hughes77}), which are listed in
Table \ref{tab:resSRP}.
\begin{table}
	\centering
	\caption{List of the first-order terms, expanding the SRP disturbing function up to first-order (e.g., \cite{Hughes77}): argument $\psi_j$, values of the coefficients $\alpha,\,\beta,\,\gamma$ and corresponding index $j$.}
	\label{tab:resSRP}
	\begin{tabular}{lrrrc} 
		\hline
		Argument $\psi_j$ & $\alpha$ & $\beta$ & $\gamma$ & index $j$\\
		\hline
		$\omega -2\lambda_S$ & 0 & 1 & $-2$ & 12 \\
		$\omega +2\lambda_S$ & 0 & $1$ & $2$ & 13 \\
		$\Omega+\omega -2\lambda_S$ & 1 & 1 & $-2$ & 14 \\
                $\Omega +\omega$ & 1 & 1 & 0 & 15 \\
                $\Omega -\omega -2\lambda_S$ & 1 & $-1$ & $-2$ & 16 \\
                $\Omega +\omega +2\lambda_S$ & 1 & $1$ & 2 & 17 \\
                \hline
	\end{tabular}
\end{table}

\begin{figure}
\centering
	\includegraphics[width=0.80\textwidth]{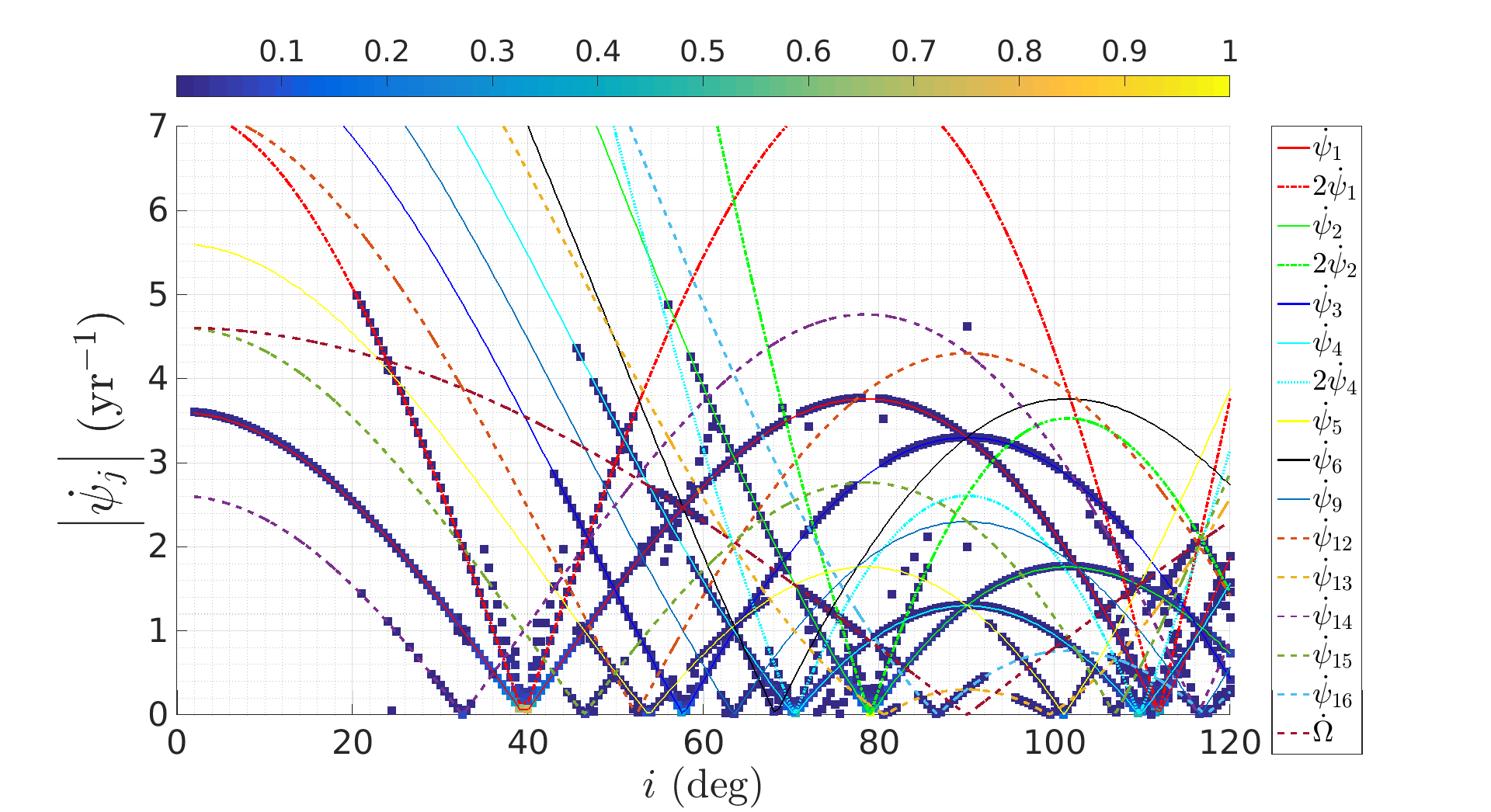}
        \includegraphics[width=0.80\textwidth]{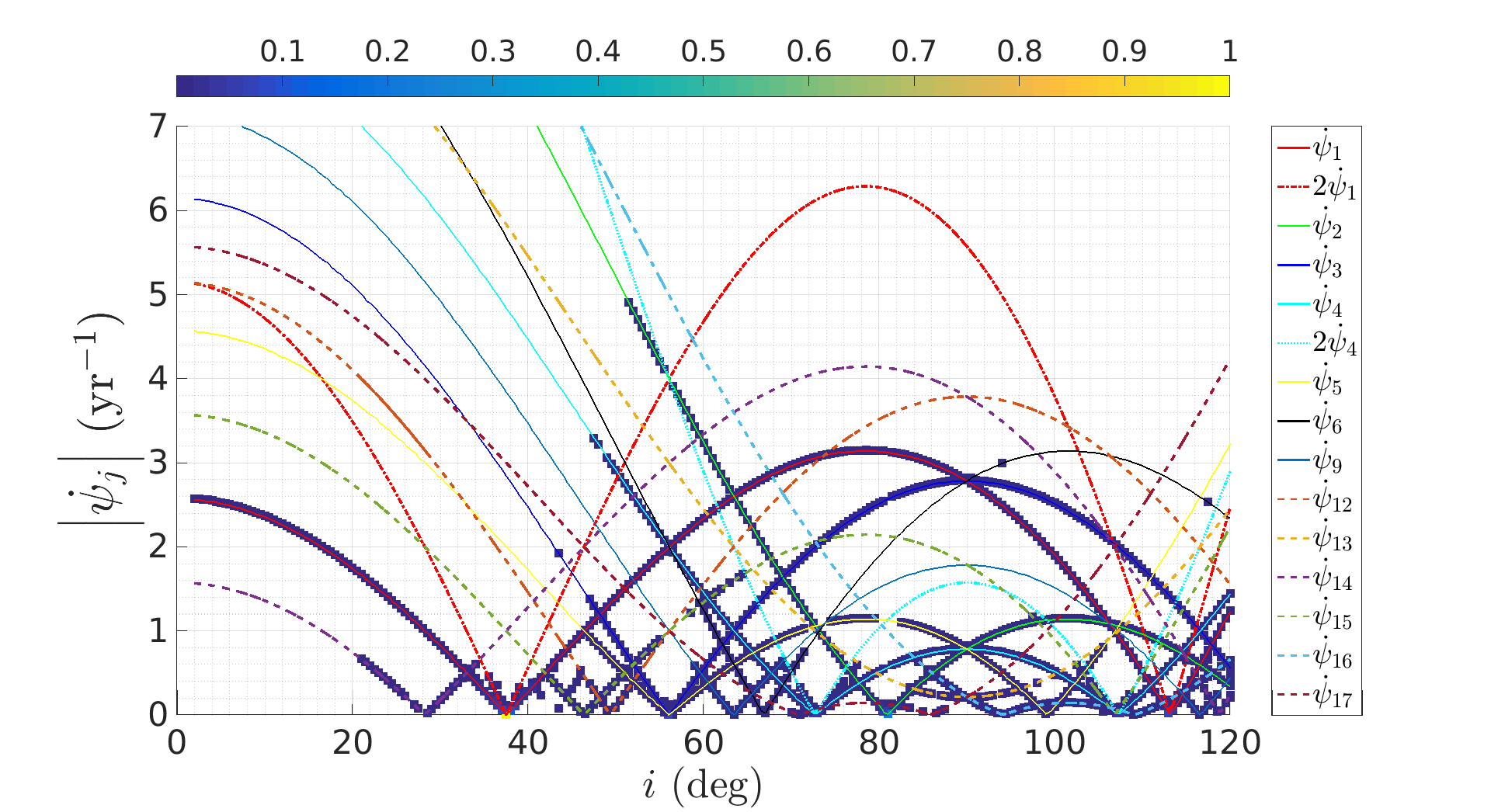}
        \caption{Frequency signatures (filled squares) detected at
          each inclination for the initial orbit at $a=7978$ km (top)
          and $a=8578$ km (bottom), for initial $e=0.001$, with
          $A/m=1$ m$^2/$kg. The $|\dot\psi_j|$ curves are those
          associated to SRP resonances, shown in Tables \ref{tab:res}
          and \ref{tab:resSRP}. The color bar refers to the relative
          amplitude of the frequency signature normalised to the
          maximum detected amplitude.}
    \label{fig:freq_modI}
\end{figure}

Following the procedure depicted in Section~\ref{sec:fc}, we
identified the main frequency signatures associated to the
eccentricity, at each initial condition available. The frequency
components detected at each inclination for the two illustrative cases
of an initial orbit at $a=7978$ km and $a=8578$ km in the case of
initial $e=0.001$, with $A/m=1$ m$^2/$kg, are shown in
Figure~\ref{fig:freq_modI}.
Each square in the plot represents a detected frequency component; the
color bar refers to the relative amplitude of the frequency
signature\footnote{The amplitude of each signature is normalised to
  the maximum detected amplitude, found in this case at the resonance
  $\dot\psi_1\simeq 0$.}, intended as the corresponding intensity peak
in the computed Fourier spectrum. Each coloured curve represents the
behaviour of the argument $|\dot\psi_j |$ as a function of the
inclination, with a cusp at the resonant inclination. As it can be
seen, the detected signatures match almost exactly the theoretical
curves. We also point out that the amplitude of the signatures
gradually grows approaching a resonant inclination. In particular,
the effects of SRP first-order terms at given inclinations, which
could be only partially inferred from the maximum eccentricity maps,
can be clearly identified in the frequency chart.

The signatures detected by means of the frequency analysis match the
bright corridors detected in the maximum eccentricity maps in the left
of Figure~\ref{contour1}. In particular, resonances 3 and 5 (see Table
\ref{tab:res}), which intersect for $a=8578$ km, can be individually
identified for $a=7978$ km both in the contour map and in the
frequency chart. Moreover, the frequency chart for $a=8578$ km shows a
signature around $i=86^{\circ}$ corresponding to the first-order
$\dot\psi_{17}$ term, which does not appear for $a=7978$ km neither in
the contour map nor in the frequency chart. Finally, the $a=7978$ km
chart shows a signature with singularity at $i=90^{\circ}$ which can
be associated to the rate of $\Omega$, appearing in the second-order
expansion of the SRP disturbing function (see, e.g., \cite{Hughes77}).

From the lifetime maps in the right panels of Figure~\ref{contour1},
we know that only in the case of resonance 1 SRP alone can drive a
reentry. Nevertheless, the maximum eccentricity maps show that in the
vicinity of a resonance a certain growth of eccentricity occurs
anyway. Thus, in the perspective of designing passive disposals and
when dealing with operational issues, it is crucial to consider the
timescale over which the eccentricity variation takes place.  With
this in mind, assessing the change in eccentricity led by a
perturbation without performing the numerical propagation, i.e., by
characterising the LEO phase space in terms of frequencies, represents
a very powerful tool.

In \cite{AlessiMNRAS}, starting from
Eq.~(\ref{dedt}) we showed that the maximum eccentricity variation
achievable due to the zero-order SRP resonance $j$ for a given initial
$(a,e,i)$ can be estimated as:
\begin{equation}
  \Delta e_j=\left |\frac{T_j(a,e,i)}{\dot\psi_j}\right | \,.
\label{eq:De}
\end{equation}
On the other side, the amplitude associated to each detected frequency
signature in the Fourier transform gives an estimate of the
eccentricity increment, as well. Both these values can be compared
with the numerically computed maximum eccentricity over 600 years: the
three estimates are expected to comply with each other. 

\begin{figure}
\centering
	\includegraphics[width=0.8\textwidth]{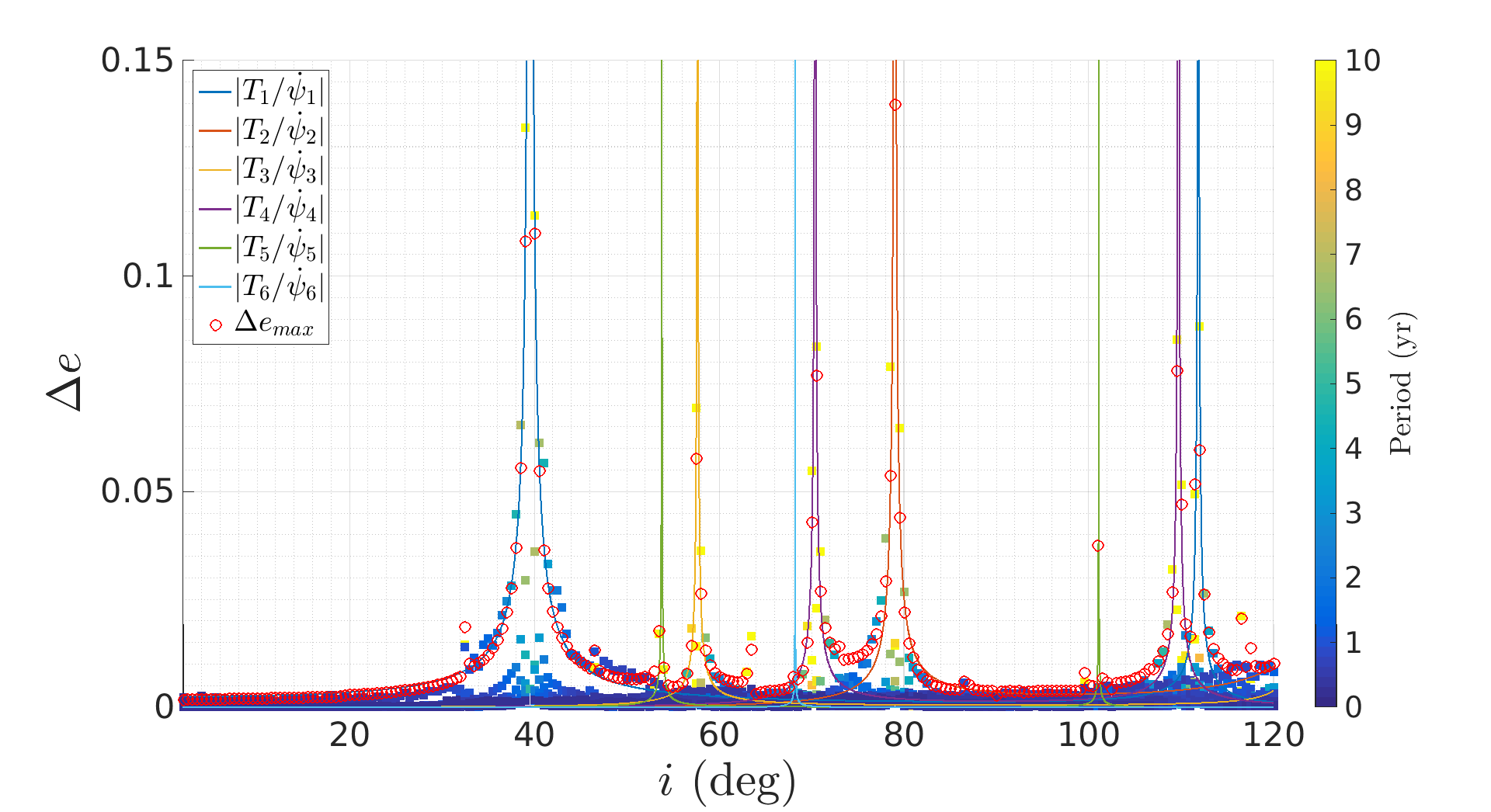}
        \caption{Theoretical amplitude $|T_j/\dot\psi_j|$ with
          $j=1,..6$ for the six zero-order SRP resonances (solid
          lines), maximum variation in eccentricity over propagation
          computed with FOP (red circles) and the frequency signatures
          detected by our analysis (filled squares; the color bar
          refers to the corresponding periodicity) as a function of
          the inclination, for the initial orbit at $a=7978$ km and
          $e=0.001$.}
    \label{Fig4}
\end{figure}

A general comparison for the initial orbit at $a=7978$ km and
$e=0.001$ is shown in Figure~\ref{Fig4}: as a function of $i$, we show
the theoretical amplitude $|T_j/\dot\psi_j|$ with $j=1,..6$ for the
six zero-order SRP resonances, the maximum variation in eccentricity,
$\Delta e_{max}$, achieved over the numerical propagation (red
circles) and the amplitude of frequency signatures detected by our
analysis (filled squares; the color bar refers to the corresponding
periodicity, i.e. the inverse of the detected frequency). A similar
example for an orbit at $a=8578$ km is shown in \cite{SchettinoIAC}.
The match is very good; moreover, we can observe that, as expected,
the brighter squares, associated to signatures with longer
periodicity, are found only in the vicinity of resonances. 

Looking at the maximum variation in eccentricity achieved during
propagation with FOP (red circles), some fainter features can be
noticed at inclinations different from those corresponding to the six
main resonances. Comparing the inclination of these signatures with
the resonant inclinations corresponding to the arguments shown in
Table \ref{tab:resSRP}, these fainter features can be associated to
the first-order terms in the expansion of the SRP disturbing function.

\begin{figure}
	\includegraphics[width=\textwidth]{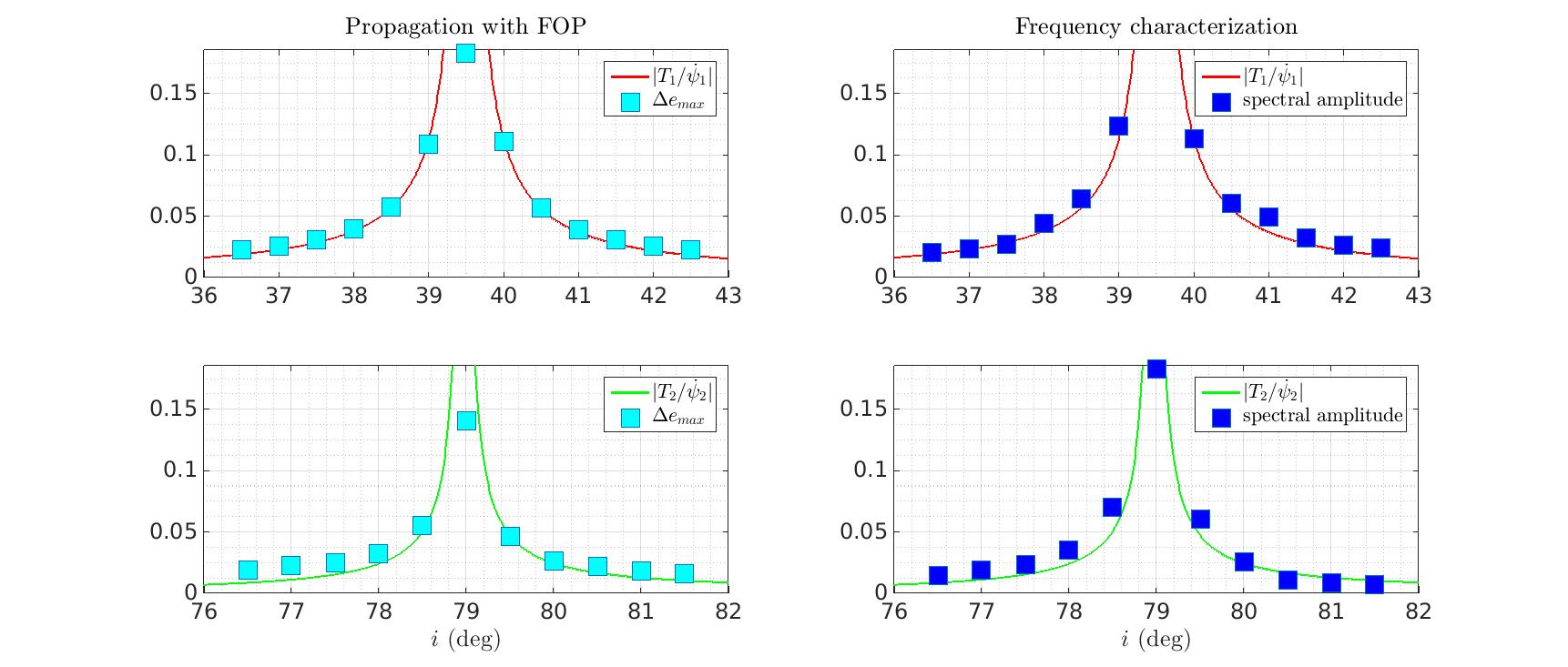}
        \caption{Comparison between theoretical amplitude
          $|T_j/\dot\psi_j|$ ($j=1$ on the top, $j=2$ on the bottom),
          maximum variation in eccentricity over propagation computed
          with FOP (left panels) and the frequency amplitudes detected
          by our analysis (right panels) as a function of the
          inclination, for the initial orbit at $a=7978$ km and
          $e=0.001$ in the case of model I.}
    \label{Fig5}
\end{figure}

In Figure~\ref{Fig5}, we show a detailed $(i,e)$ zoom around the two
main resonances found at this altitude: $\dot\psi_1$ corresponding to
$i\sim 40^{\circ}$ and $\dot\psi_2$ in the vicinity of
$i\sim 80^{\circ}$. The maximum eccentricity displayed on the $y-$axis
corresponds to the eccentricity needed to lower the perigee down to
120 km, $e_{120\textrm{km}}=0.185$.
Both the squares corresponding to the numerical maximum eccentricity
(left panels) and the amplitude of the frequency signatures (right
panels) lie on the theoretical curves for $|T_1/\dot\psi_1|$ and
$|T_2/\dot\psi_2|$. This further confirms that the three quantities
(theoretical amplitude, numerical maximum eccentricity and amplitude
of the frequency signature) provide the same information, thus one can
be adopted in place of the other.

\begin{figure}
\centering
	\includegraphics[width=0.65\textwidth]{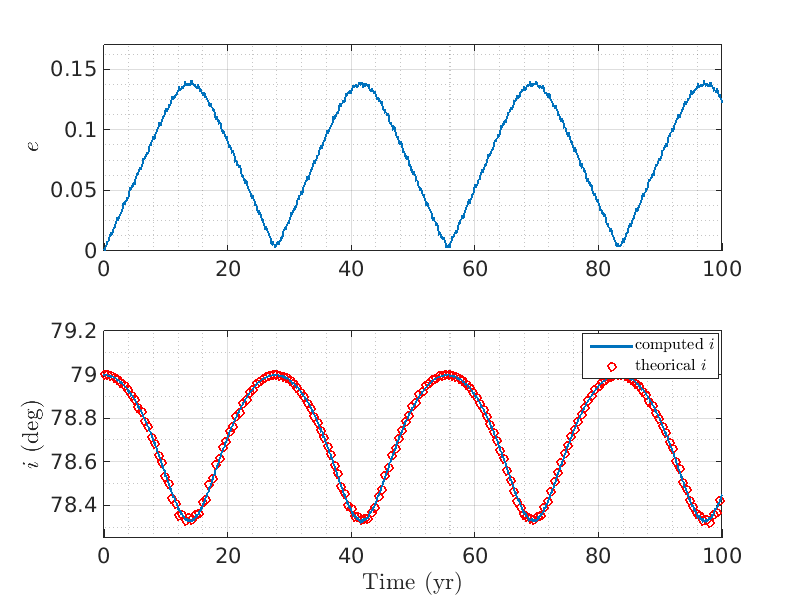}
        \caption{Eccentricity (top) and inclination (bottom) evolution
          over 100 years for initial condition $a=7978$ km, $e=0.001$,
          $i=79^{\circ}$ in the case of model I. The inclination
          computed by propagation (blue line) is compared with the
        theoretical inclination (red circles) derived from
        Eq.~(\ref{eq:i}).}
    \label{ie79}
\end{figure}

We can notice, however, in the bottom panel on the left of
Figure~\ref{Fig5}, a disagreement between the theory and the numerical
propagation: according to the theory, the maximum eccentricity
variation for initial $i=79^{\circ}$ should be sufficient to lead to
reenter, while the $\Delta e_{max}$ computed with FOP turns out to be
lower than $e_{120\textrm{km}}$. The explanation for such a behaviour
is that during the propagation also the inclination experiences a
variation which moves the object away from the resonance, making the
SRP perturbation less effective. In Figure~\ref{ie79} we show the
evolution of $e$ and $i$ over 100 years for the initial condition
$a=7978$ km, $e=0.001$, $i=79^{\circ}$.
Both eccentricity and inclination show a periodicity of about 28 years
but they are out of phase: the eccentricity starts to grow led by the
SRP perturbation; at the same time, the inclination starts to decrease
so that when the eccentricity reaches the maximum value
$e_{max}=0.14$, the inclination is at its minimum,
$i_{min}=78.3^{\circ}$, where, as can be inferred from
Figure~\ref{Fig5}, the perturbation due to the resonant term $\psi_2$
is no longer effective in driving the reentry.

This fact shows that the rate of $i$ should be taken into account to
provide a full description of this case based on the dynamics. It is
beyond the scope of this work to provide a full description on this scenario
based on the dynamical systems theory, but we can provide a basic tool
to obtain an a priori indication on whether the orbit will exit from
the resonance domain before achieving a reentry.

\begin{figure}
\centering
	\includegraphics[width=0.7\textwidth]{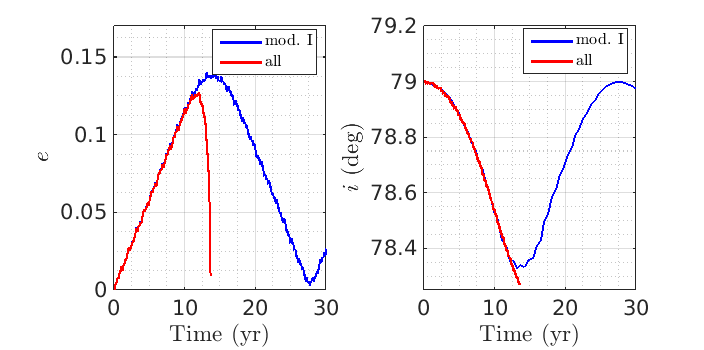}
        \caption{Comparison of the eccentricity (left) and inclination (right) evolution over 30
          years, computed by propagation assuming model I (blue curve) and including
          all the perturbations provided by FOP (red curve). The
          initial orbit is for both cases: $a=7978$ km, $e=0.001$,
          $i=79^{\circ}$.}
    \label{ie79_tot}
\end{figure}

In Figure~\ref{ie79}, in the panel showing the evolution of the
inclination, it is also displayed the behaviour predicted by the
theory developed in \cite{D16} for lunisolar gravitational
resonances, which can be applied also in the case of SRP, as shown in
\cite{AlessiMNRAS}. In particular, it is demonstrated that there
exists an integral of motion, corresponding to
\begin{equation}\label{eq:const}
(\beta\cos{i}-\alpha)\sqrt{\mu a(1-e^2)}=\textrm{constant},
\end{equation}
where $\alpha, \beta$ are as defined in Eq.~(\ref{eq:psi}). In other
words, assuming that the motion of the spacecraft is governed only by
the Earth's monopole, the Earth's oblateness and the solar radiation
pressure, at any instant we can recover the inclination value from
\begin{equation}\label{eq:i}
i=\pm\arccos{\left(\frac{\textrm{constant}}{\beta\sqrt{\mu a(1-e^2)}}+\frac{\alpha}{\beta}\right)},
\end{equation}
where the constant can be obtained by evaluating Eq.~(\ref{eq:const})
at the initial epoch. For completeness, in Figure~\ref{ie79_tot} we
show a comparison over 30 years of the eccentricity and inclination
evolution computed by propagation assuming model I (blue curve)
with the behaviour obtained by assuming the complete dynamical model
(red curve), which includes all the perturbations provided by FOP. The
initial orbit is the same as in Figure~\ref{ie79}. We can observe that
the two models predict the same behaviour, except that, in the second
case, the reentry is ensured (in 13.6 years) by the atmospheric drag.

\begin{figure}
\centering
	\includegraphics[width=0.45\textwidth]{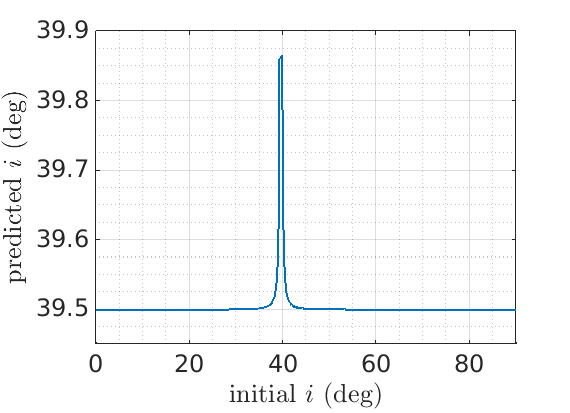} 
        \includegraphics[width=0.45\textwidth]{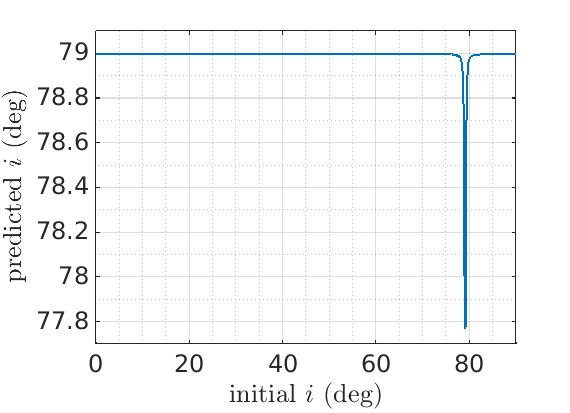}
        \caption{Predicted inclination variation as a function of the initial inclination, assuming model I, $a=7978$ km, $e=0.001$. Left: resonance 1. Right: resonance 2.}
    \label{fig:ivariation}
\end{figure}

In Figure~\ref{fig:ivariation}, we show the behaviour predicted for the
inclination by Eq.~(\ref{eq:i}), by assuming a maximum variation in
eccentricity as in Eq.~(\ref{eq:De}), for resonances $1$ and $2$. We can
notice that in the first case, when we consider an initial inclination
in the resonance domain, the variation is not relevant if compared with
the curves in the top panel of Figure~\ref{Fig5}). In the second case,
the variation is instead important, of about $1^{\circ}$ and moves the
dynamics towards the edges of the interval where the resonance is
effective (compare with the curves in the bottom panel of
Figure~\ref{Fig5}).

\begin{figure}
\centering
	\includegraphics[width=0.6\textwidth]{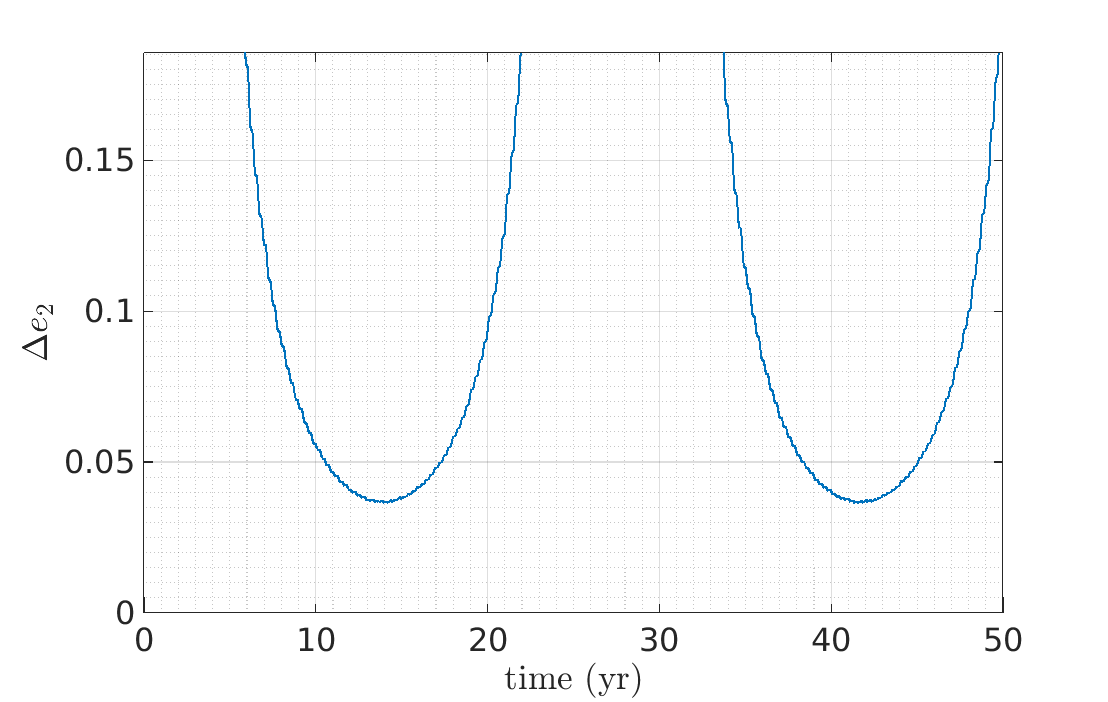}
        \caption{Evolution of $\Delta e_2=T_2/|\dot\psi_2|$ over 60
          years, computed by means of Eq.~(\ref{eq:De}) on the $e$ and
          $i$ values obtained by propagation with FOP in case of model
          I, for an initial $a=7978$ km.}
    \label{fig:T2_e_i}
\end{figure}

The above discussion showed that the assumption that $T_j$ is a
function of the initial values of eccentricity and inclination may
provide a misleading information. Figure~\ref{fig:T2_e_i} shows the
evolution of $\Delta e_2=T_2(a,e,i)/|\dot\psi_2|$, according to
Eq.~(\ref{eq:De}), assuming the values of eccentricity and inclination
computed at each given time by propagation with FOP, in case of model
I, for initial $a=7978$ km, $e=0.001$, $i=79^{\circ}$. The $y-$axis
upper limit corresponds to a perigee altitude of 120 km. As it can be
seen, for the initial value of $e$ and $i$, the growth of eccentricity
$\Delta e_2$ is such that the reentry driven by resonance 2 is
feasible (the curve is not visible in the figure because it is higher
than the eccentricity required to reentry). On the contrary, after
only 5 years, the inclination has moved from its initial value
(compare with Figure~\ref{ie79}) enough that the corresponding growth
in eccentricity due to resonance 2 alone is no more capable to assure
the reentry.

\begin{figure}
	\includegraphics[width=\textwidth]{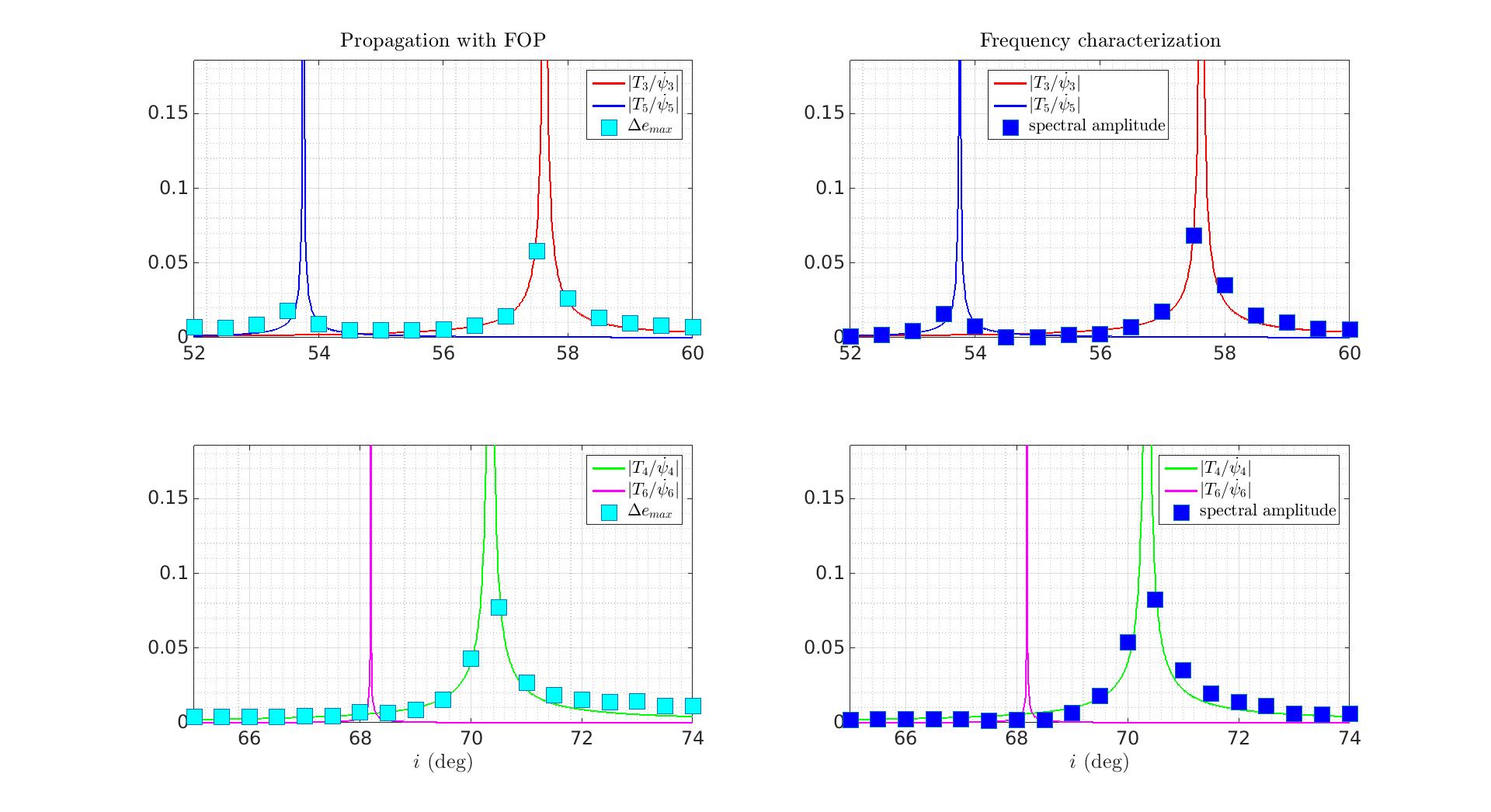}
        \caption{Comparison between theoretical amplitude
          $|T_j/\dot\psi_j|$ ($j=3,4,5,6$), maximum variation in eccentricity over
          propagation computed with FOP (left panels) and the frequency
          amplitudes detected by our analysis (right panels) as a
          function of the inclination, for the initial orbit at
          $a=7978$ km and $e=0.001$ in tha case of model I.}
    \label{Fig5bis}
\end{figure}

Finally, similarly to Figure~\ref{Fig5}, the comparison between
theoretical amplitude, maximum variation in eccentricity computed with
FOP and amplitude of the frequency signatures for $a=7978$ km and
$e=0.001$ in the cases of resonances $3,\,4,\,5,\,6$ due to SRP is
shown in Figure~\ref{Fig5bis}. Also in these cases the agreement is
noticeable.

\subsection{Model II}

\subsubsection{Analysis in the time domain}
\label{modII_time}

Model II is particularly suitable to study the perturbation on
eccentricity due to lunisolar effects and high-degree terms in
geopotential, since SRP has been removed in this case. The effective
area-to-mass ratio of the object does not play a role in driving the
dynamics, contrary to the case of the previous model, thus we assume
$A/m=0.012$ m$^2/$kg for simulations.

\begin{figure}
  \begin{center}   
  \includegraphics[width=0.48\textwidth]{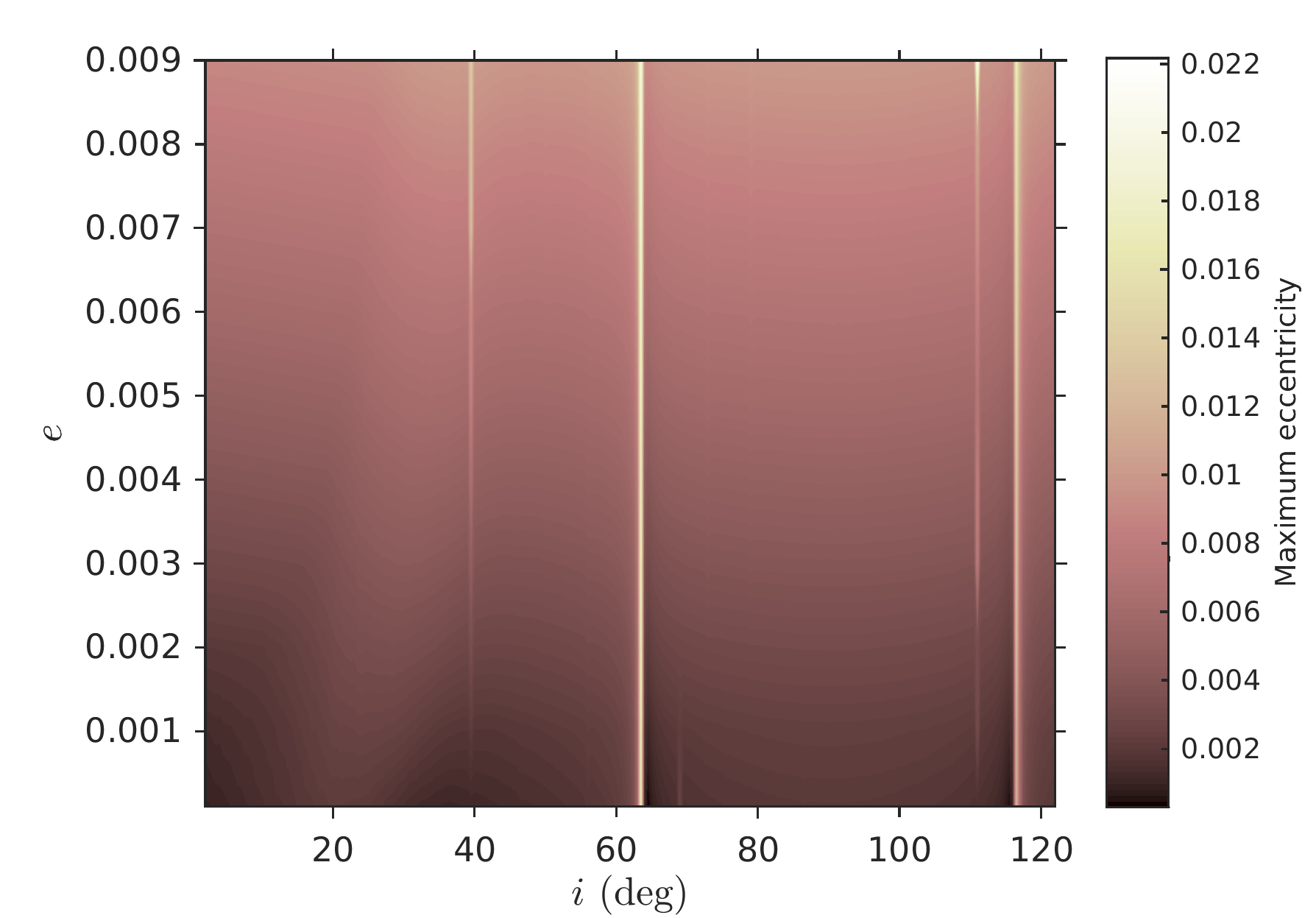}
  \includegraphics[width=0.48\textwidth]{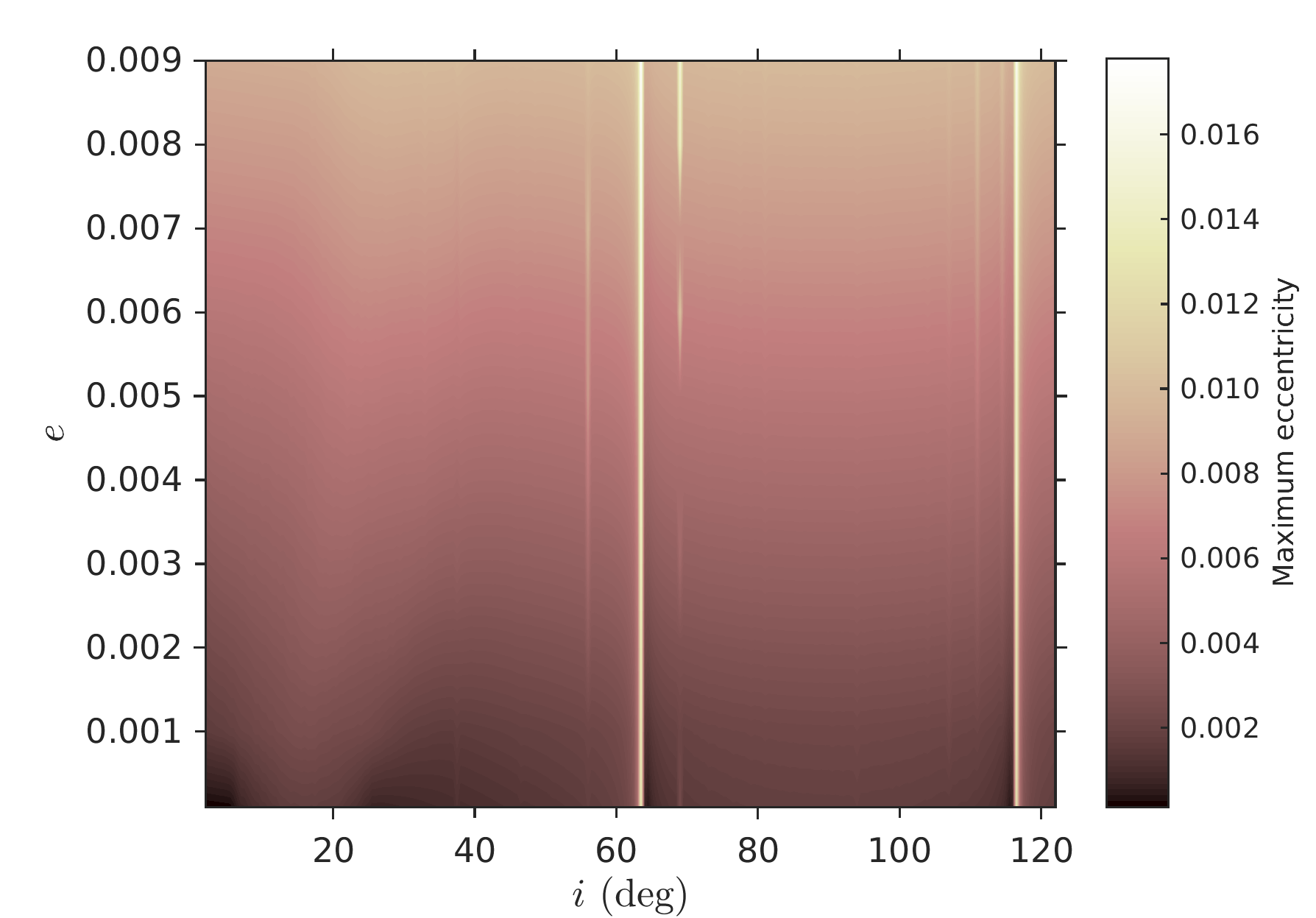}
 \caption{Maximum eccentricity as a function of
   the initial inclination at steps of $\Delta i=0.5^{\circ}$ and $e$
   at steps of $\Delta e=0.001$, assuming model II and $A/m=0.012$
   m$^2/$kg, for the initial orbits at $a=7978\,$km (left) and
   $a=8578\,$km (right), with $\Omega=\omega=0^{\circ}$ and initial
   epoch 21 June 2020.}
      \label{contour2} 
\end{center}
\end{figure}

In analogy to the left panels of Figure~\ref{contour1}, Figure~\ref{contour2}
shows the maximum eccentricity as a function of the initial
inclination and eccentricity for an orbit at $a=7978$ km (left) and
$a=8578$ km (right), respectively. In this case, we do not show the
corresponding lifetime maps: at these altitudes and for quasi-circular
orbits the maps would result blank since neither lunisolar
perturbations nor high-degree terms of geopotential are capable to
induce a growth of eccentricity such that the perigee is lowered down
to altitudes where drag becomes effective. The synergic effect of drag
and other perturbations can be possibly exploited at these altitudes
only for initial eccentricities higher than 0.1\footnote{Contour maps
  similar to Figure~\ref{contour2} including eccentricities up to 0.28
  can be found on the project website.}. The most evident signatures
in the maximum eccentricity maps are those at
$i=63.4^{\circ},116.6^{\circ}$, also known as \textit{critical
  inclinations} (e.g., \cite{Beutler}), which corresponds to the
condition $\dot\omega=0$ (resonance 9 in Table~\ref{tab:res}).

\begin{figure}
  \begin{center}   
  \includegraphics[width=0.32\textwidth]{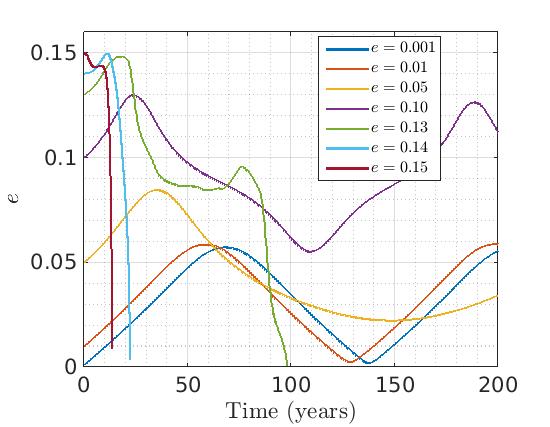}
       \includegraphics[width=0.32\textwidth]{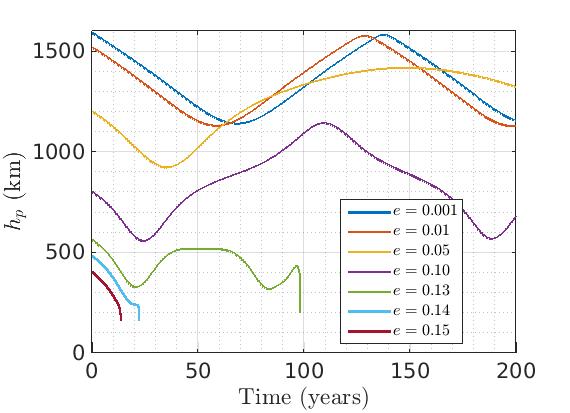}
  \includegraphics[width=0.32\textwidth]{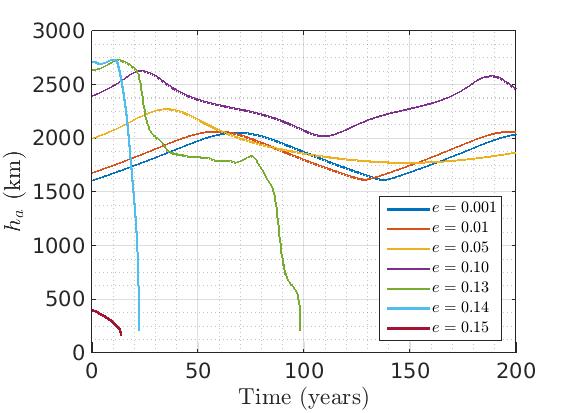}
  \includegraphics[width=0.32\textwidth]{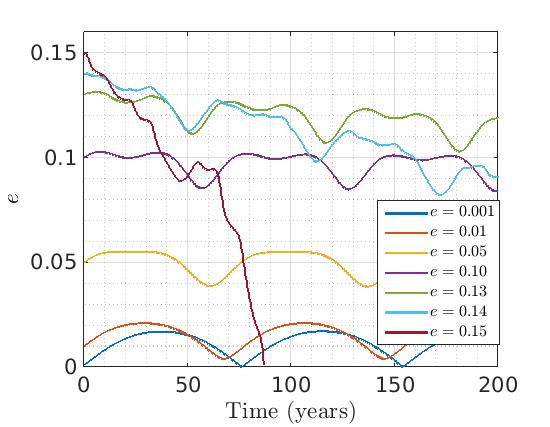}
       \includegraphics[width=0.32\textwidth]{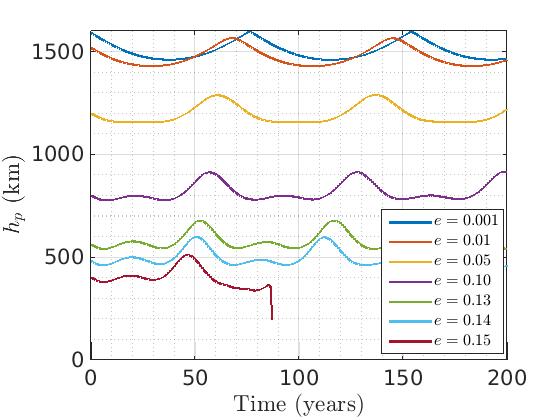}
  \includegraphics[width=0.32\textwidth]{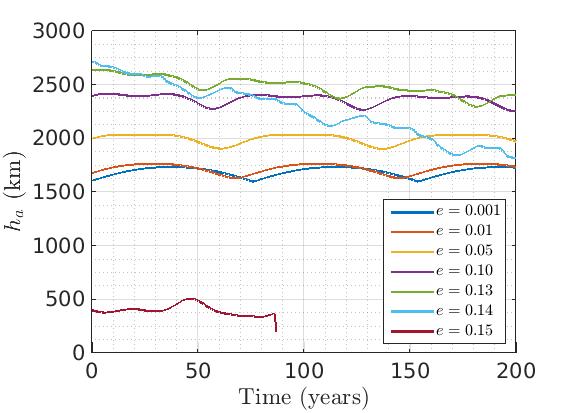}
\caption{Time evolution of eccentricity (left), perigee altitude
  (middle) and apogee altitude (right) over 200 years of propagation
  with FOP, for initial $a=7978$ km and $i=63.4^{\circ}$ (top),
  $i=63.5^{\circ}$ (bottom), for 7 different initial eccentricities:
  $e=0.001,0.01,0.05,0.10,0.13,0.14,0.15$ in the case of model II,
  assuming as initial epoch 21 June 2020.}
      \label{zoom_1600} 
\end{center}
\end{figure}

Figure~\ref{zoom_1600} depicts the time evolution of different orbits
with initial $a=7978$ km, considering two different initial
inclinations: $i=63.4^{\circ}$ (top), which corresponds exactly to the
resonant inclination for the condition $\dot\omega=0$, and
$i=63.5^{\circ}$ (bottom), i.e., only $0.1^{\circ}$ degrees next to
the resonant value. The initial eccentricity varies from $0.001$ to
$0.15$: on the left, we show the evolution of eccentricity over 200
years, in the middle, the pericenter altitude and on the right, the
apocenter altitude. As it can be seen, the behaviour is different if
the initial inclination corresponds exactly to the resonant value or
not. Up to initial $e=0.1$, for both inclinations the eccentricity
does not experience a sufficient growth to lower the perigee in order
to reenter. Indeed, for the case of an initial quasi-circular orbit
($e=0.001$), at resonance the perigee lowers only by 70 km after 10
years and 177 km after 25 years, while for $i=63.5^{\circ}$ the
decrease of the perigee is 58 km after 10 years and 115 km after 25
years.

At resonance we can observe that the characteristic period of the
eccentricity evolution is clearly longer than in the
neighborhood of the resonance. For example, for $i=63.4^{\circ}$ and
$e=0.001$, the eccentricity shows a period of 137 years, while for
$i=63.5^{\circ}$ it reduces to 76 years.  For higher eccentricities,
such as $e=0.13$ and $e=0.14$, at $i=63.4^{\circ}$ the initial growth
of eccentricity induced by the perturbation lowers the perigee down to
an altitude where atmospheric drag becomes effective. Conversely, for
$i=63.5^{\circ}$ the apogee starts to lower while the perigee is not
low enough for drag to be effective in less than 200 years. Finally,
for $e=0.15$ the perigee is low enough that reentry is feasible at
both initial inclinations thanks to the atmospheric drag.

Looking at Figure~\ref{contour2}, other fainter signatures at given
inclinations can be recognised:
\begin{itemize}
\item at $i\simeq 40^{\circ},113^{\circ}$, in the $a=7978$ km panel,
  corresponding to the well-known \textit{evection resonance}
  (e.g., \cite{Brouwer})
  $\dot\psi_8=2(\dot\Omega+\dot\omega-\dot\lambda_S)\simeq 0$;
\item at $i\simeq 56^{\circ}$, visible in the $a=8578$ km panel,
  corresponding to the condition
  $\dot\psi_{10}=\dot\Omega+2\dot\omega\simeq 0$;
\item at $i\simeq 70^{\circ}$, clearly recognisable at $a=8578$ km
  while distinguishable only for very low eccentricities at $a=7978$
  km, which corresponds to the resonant condition
  $\dot\Omega-2\dot\omega\simeq 0$, as will be discussed in
  Sect.~\ref{sec:3.modII}.
\end{itemize}

\subsubsection{Analysis in the frequency domain}
\label{sec:3.modII}

\begin{figure}
\centering
	\includegraphics[width=0.80\textwidth]{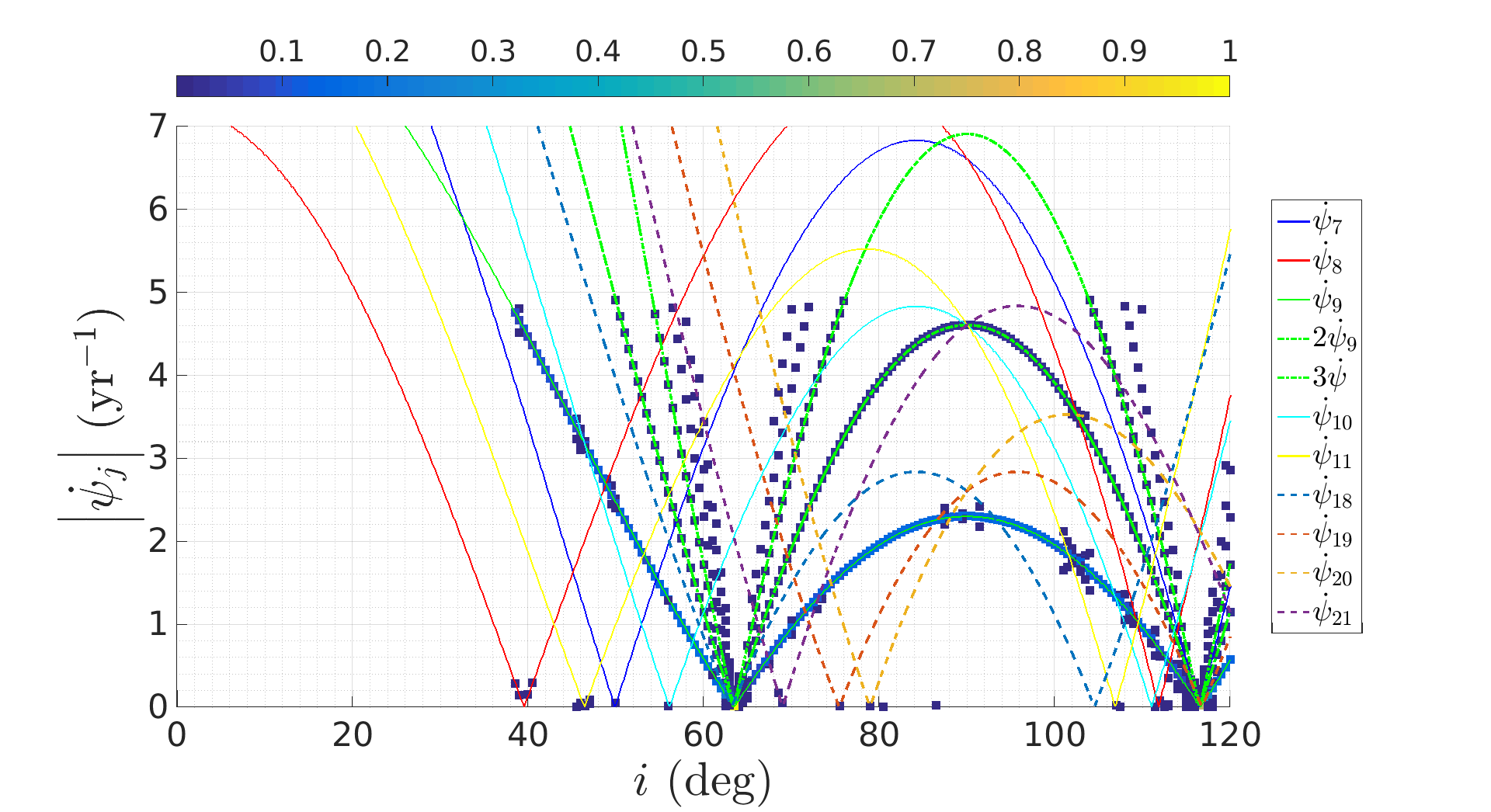}
        \includegraphics[width=0.80\textwidth]{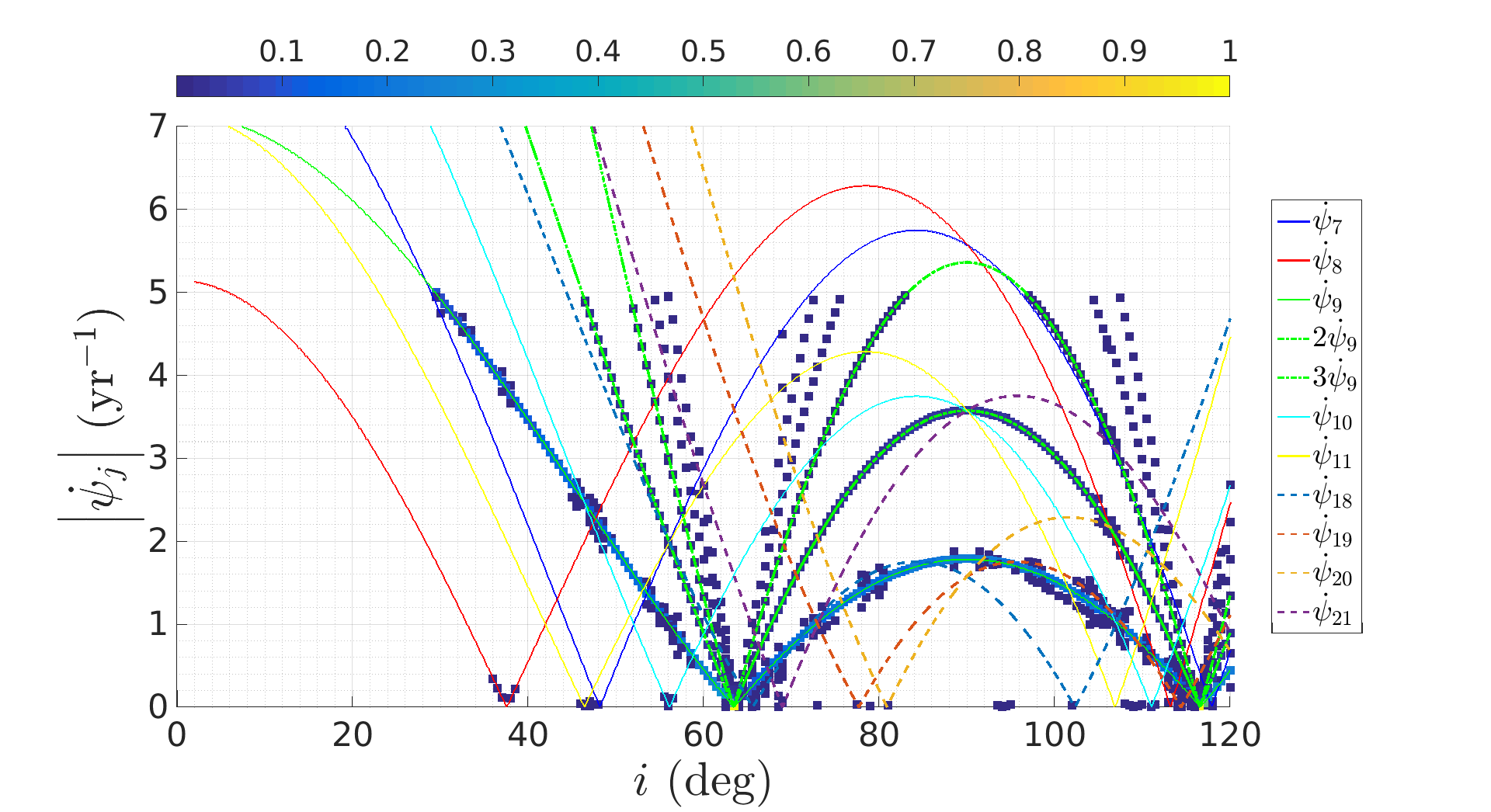}
        \caption{Frequency signatures (filled squares) detected at
          each inclination for the initial orbits at
          $a=7978$ km (top) and $a=8578$ km (bottom),
            assuming initial $e=0.001$, with $A/m=0.012$
          m$^2/$kg. The $|\dot\psi_j|$ curves are those associated to
          lunisolar resonances, shown in Tables \ref{tab:res} and
          \ref{tab:resLS}. The color bar refers to the relative
          amplitude of the frequency signature normalised to the
          maximum detected amplitude.}
    \label{fig:freq_modII}
\end{figure}

The frequency components detected at each inclination for the initial
orbits at $a=7978$ km and $a=8578$ km, assuming initial $e=0.001$,
with $A/m=0.012$ m$^2/$kg are shown in Figure~\ref{fig:freq_modII},
where each frequency signature corresponds to a filled square and the
color bar refers to the relative amplitude found in the Fourier
spectrum.

The solid curves in the figure represent the resonant arguments
$\psi_j$, with $j=7,..11$, associated to solar gravitational and
lunisolar perturbations, shown in Table~\ref{tab:res}; the dashed
curves refer, instead, to fainter, but still detectable, signatures
listed in Table~\ref{tab:resLS}. They correspond to the arguments
$\psi_j$ with $j=18,..20$ associated to singly-averaged solar
gravitational resonances, and to the argument $\psi_{21}$ associated
to doubly-averaged lunisolar perturbations \cite{Hughes80}.

\begin{figure}
\centering
	\includegraphics[width=0.80\textwidth]{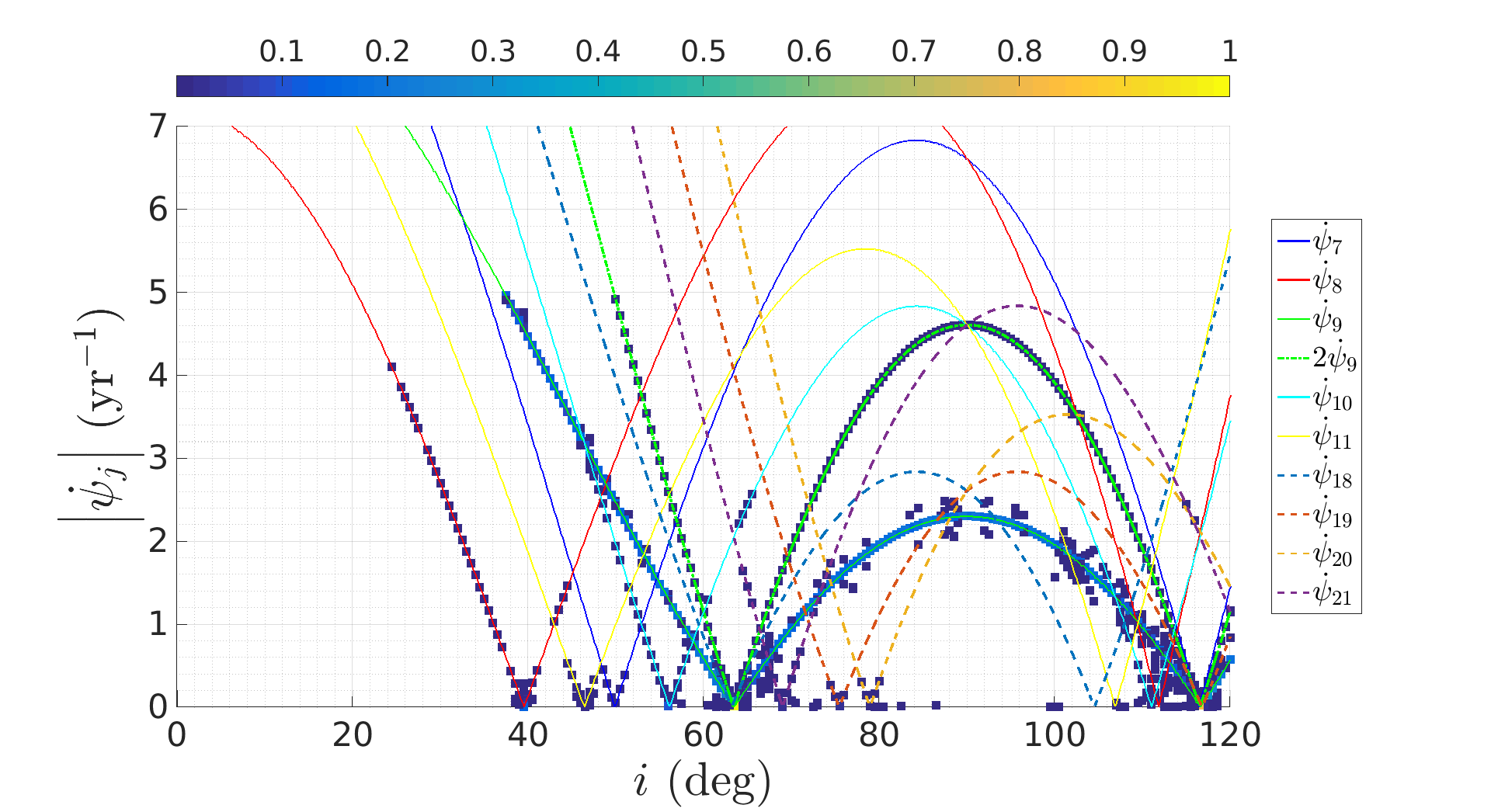}
        \includegraphics[width=0.80\textwidth]{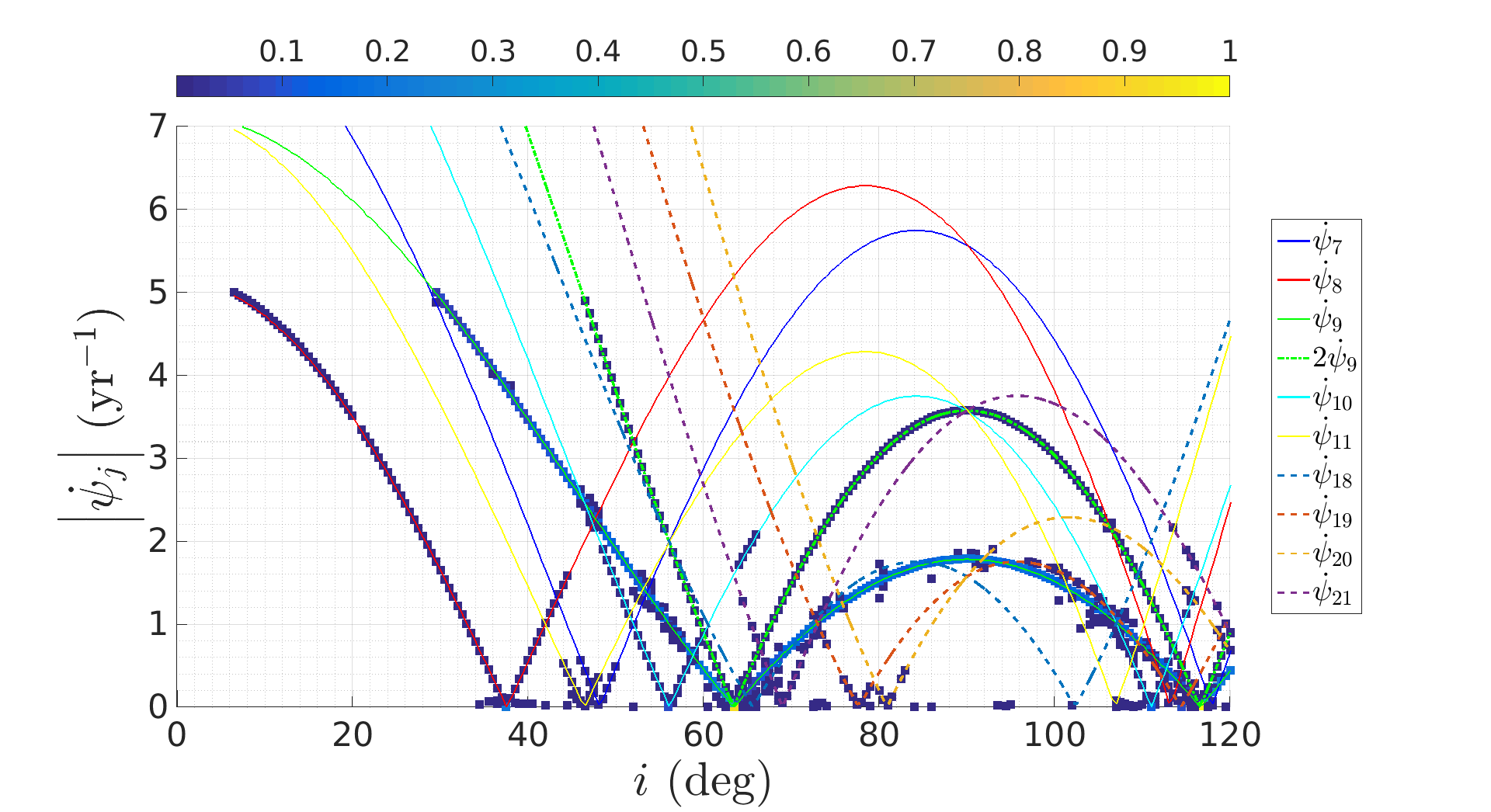}
        \caption{Frequency signatures (filled squares) detected at
          each inclination for the initial orbits at
          $a=7978$ km (top) and $a=8578$ km (bottom),
          assuming initial $e=0.01$ and $A/m=0.012$ m$^2/$kg. The
          $|\dot\psi_j|$ curves are those associated to lunisolar
          resonances, shown in Tables \ref{tab:res} and
          \ref{tab:resLS}. The color bar refers to the relative
          amplitude of the frequency signature normalised to the
          maximum detected amplitude.}
    \label{fig:freq_modII_2}
\end{figure}

The main signature in both frequency charts is the one at
$i=63.5^{\circ}$ associated to resonance 9, which corresponds also to
the brightest corridor in the eccentricity contour maps of
Figure~\ref{contour2}. Concerning resonance 8, around $i=40^{\circ}$,
the contour maps showed that it is not expected to be relevant for
$e=0.001$, while it becomes more important for more eccentric
orbits. Indeed, it is only partially detectable in the $e=0.001$
frequency charts of Figure~\ref{fig:freq_modII}, while its role
becomes more evident in the frequency charts of
Figure~\ref{fig:freq_modII_2}, which correspond to the same initial
orbits of Figure~\ref{fig:freq_modII} but with $e=0.01$. Comparing the
frequency charts corresponding to the two values of eccentricity, we
can notice also that resonances $7\,,8\,,11$ and the higher order
resonances shown in Table~\ref{tab:resLS} are only partially
detectable in the $e=0.001$ frequency charts, while they are clearly
recognizable for the $e=0.01$ ones.
In particular, the signature due to the $\dot\psi_{21}$ term is
clearly visible in the $a=8578$ km maximum eccentricity map of
Figure~\ref{contour2} as the bright corridor at $i=69^{\circ}$, and it
appears also in the corresponding frequency chart.
\begin{table}
	\centering
	\caption{List of the other detected resonances due to lunisolar perturbations \cite{Hughes80}: argument $\psi_j$, values of the coefficients $\alpha,\,\beta,\,\gamma$ and corresponding index $j$.}
	\label{tab:resLS}
	\begin{tabular}{lrrrc} 
		\hline
		Argument $\psi_j$ & $\alpha$ & $\beta$ & $\gamma$ & index $j$\\
		\hline
		$\Omega +2\omega +2\lambda_S$ & 1 & 2 & 2 & 18 \\
		$\Omega -2\omega -2\lambda_S$ & 1 & $-2$ & $-2$ & 19 \\
		$2\Omega-2\omega -2\lambda_S$ & 2 & $-2$ & $-2$ & 20 \\
                $\Omega-2\omega $ & 1 & $-2$ & 0 & 21 \\
                \hline
	\end{tabular}
\end{table}

Figure~\ref{contour2} showed that the growth of eccentricity that can
be reached thanks to high degree zonal harmonics and/or lunisolar
perturbations, for the initial eccentricities considered, is, at most,
one order of magnitude less than exploiting SRP in the case of an area
augmentation device.

\begin{figure}
\centering
	\includegraphics[width=0.6\textwidth]{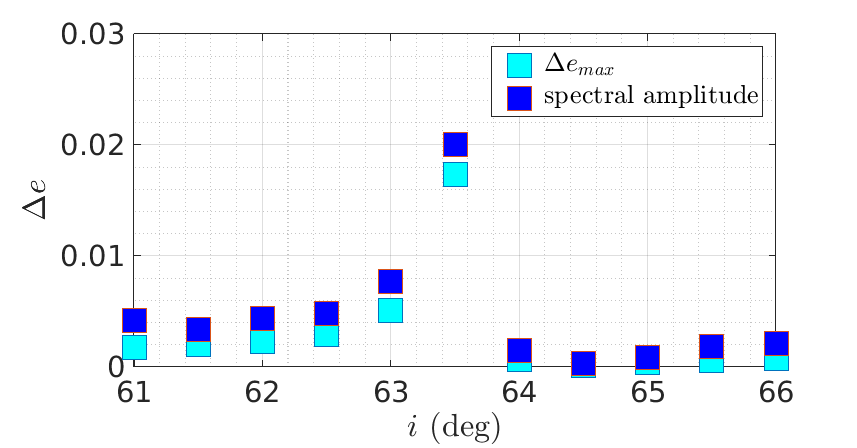}
        \caption{Comparison between the maximum variation in
          eccentricity over propagation computed with FOP (cyan
          squares) and amplitude of the frequency signatures detected
          by our analysis (blue squares) in the case of resonance 9,
          for the initial orbit at $a=7978$ km and $e=0.001$.}
    \label{fig:Fig8}
\end{figure}

\begin{figure}
\centering
	\includegraphics[width=0.7\textwidth]{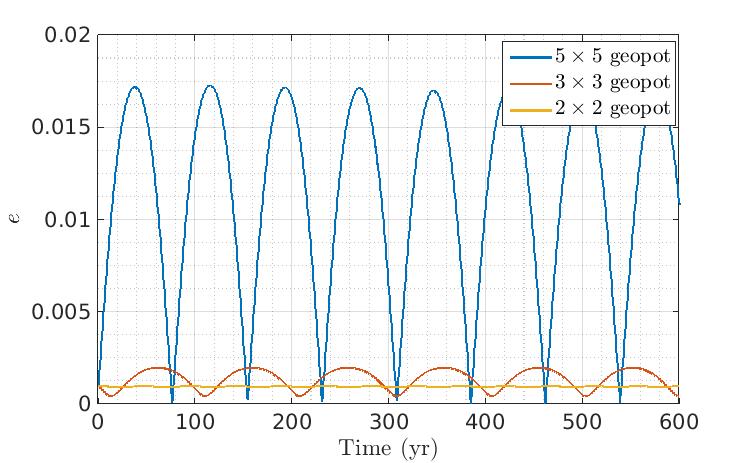}
        \caption{Evolution of $e$ for initial $a=7978$ km and
          $i=63.5^{\circ}$ for three different models, all including
          drag and lunisolar perturbations: (I) $5\times 5$
          geopotential, (II) $3\times 3$ geopotential, (III)
          $2\times 2$ geopotential.}
    \label{e_geopot}
\end{figure}

The most favourable case is found in proximity of resonance 9
($\dot\omega\simeq 0$), where $\Delta e_{max}\simeq 0.02$ can be
achieved. As already noticed, the frequency analysis shown in
Figure~\ref{fig:freq_modII} confirms this finding for both altitudes:
the main signature appears at $i=63.5^{\circ}$, corresponding to the
cusp of the $|\dot\psi_9|$ curve. Figure~\ref{fig:Fig8} compares the
behaviour of the numerical maximum eccentricity over propagation (cyan
squares) and the amplitude found through the frequency analysis (blue
squares) around $i=63.5^{\circ}$ for an initial orbit with $a=7978$ km
and $e=0.001$. As for the case of model I, there is a very good match
between the two quantities. We can notice that the growth of
eccentricity at $i=63.5^{\circ}$ is mainly due to the perturbing
effect of $J_5$. Indeed, if we consider only a $3\times 3$
geopotential instead of $5\times 5$, the increment of eccentricity
decreases from $\Delta e_{5\times 5}=0.017$ to
$\Delta e_{3\times 3}=0.002$, while if only lunisolar perturbations
and $2\times 2$ geopotential are included in the dynamical model, the
eccentricity does not experience any variation at this inclination.
These results are shown in Figure~\ref{e_geopot}, which displays the
evolution of eccentricity for initial $a=7978$ km and $i=63.5^{\circ}$
for three different models, all including drag and lunisolar
perturbations: (i) $5\times 5$ geopotential, (ii) $3\times 3$
geopotential, (iii) $2\times 2$ geopotential.

Although at high altitudes in LEO the growth of eccentricity induced
by geopotential or lunisolar perturbations is not capable to drive the
reentry, the variation in $e$ can be, anyway, not negligible. Indeed,
the perigee and apogee of the orbit can experience an oscillation which
should be taken into account if we are dealing with issues as the
stability of an operational orbit. This happens, for example, in the
considered case of initial $a=7978$ km and $e=0.001$ and assuming a
$5\times 5$ geopotential as in model II-(i): the perigee undergoes a
76 years periodic evolution with a maximum oscillation of 130 km; after 10
years it experiences a variation of 55 km, while as much as 115 km
after 25 years.

\section{Conclusions}
\label{sec:concl}

In this paper we studied the evolution of the eccentricity of a large set of
orbits both in the time and frequency domains, deepening the work
already presented by the authors in \cite{AlessiCMDA,AlessiMNRAS}.

First, we considered the role of SRP in driving the dynamics for an
object equipped with an area augmentation device. We found that, for
quasi-circular orbits, SRP can be exploited, possibly in concurrence
with the atmospheric drag, to lead the disposal within 25 years, but
only if the initial orbital inclination is close enough to the
resonant inclinations associated with the condition
$\dot\psi=\dot\Omega\pm\dot\omega-\dot\lambda_S\simeq 0$ (resonances 1
and 2). In the vicinity of the other zero-order resonances (indexed
from 3 to 6), but also in correspondence of the first-order SRP
resonances (indexed from 12 to 17), a growth of eccentricity due to
SRP takes place in any case but over longer time scales, of the order
of tens to hundreds of years. Although this variation of eccentricity
cannot be exploited for disposal, it needs to be taken into account
for operational purposes in the perspective of identifying long-term
stable orbits within LEO.

Moreover, in \cite{AlessiMNRAS} we presented a simplified theory to
analytically evaluate the growth of eccentricity induced by the six
main SRP resonances. Here, we showed that the assumption to consider
the variation of eccentricity only as a function of the initial
$(e,i)$ state could be coarse and that, for given initial orbits, also
the role of the variation of inclination over time should be
considered, to give a coherent picture of the dynamics.

Then, we focused on the role of lunisolar perturbations and high
degree zonal harmonics. In this case, the growth of eccentricity
induced by the perturbations does not cause a lowering of the perigee
leading to a reentry, in the case of quasi-circular orbits. In
particular, we analysed the case of the well-known critical
inclination, corresponding to the resonant condition
$\dot\omega\simeq 0$, for an initial quasi-circular orbit at $a=7978$
km. We verified that the computed growth of eccentricity of about 2
orders of magnitude after 40 years is mainly due to the $J_5$
perturbation, confirming the results found in \cite{AlessiCMDA}.

\paragraph{Acknowledgements}

This work is funded through the European Commission Horizon 2020,
Framework Programme for Research and Innovation (2014-2020), under the
ReDSHIFT project (grant agreement n$^{\circ}$ 687500).

\end{document}